\documentclass[12pt]{article}
\pdfoutput=1
\usepackage{epsf,amsfonts,amssymb,epsfig,amsmath,mathtools,graphics,slashed}
\usepackage{hyperref,hep,graphicx,subfig}

\makeatletter

\newcommand{\Rmnum}[1]{\expandafter\@slowromancap\romannumeral #1@}
\makeatother

\makeatletter
\newcommand*{\Rmn}[1]{\expandafter\@slowromancap\romannumeral #1@}
\makeatother

\newcommand{\adef}{\alpha_0}
\newcommand{\cdef}{\zeta_2}

\newcommand{\nn}{\notag \\}

\begin{document}

\makeatletter
\renewcommand{\theequation}{\thesection.\arabic{equation}}
\@addtoreset{equation}{section}
\makeatother

\baselineskip 18pt

\begin{titlepage}

\vfill

\begin{flushright}
Imperial/TP/2014/JG/05\\
DCPT-14/73\\
ICCUB-14-067\\
\end{flushright}

\vfill

\begin{center}
   \baselineskip=16pt
   {\Large\bf Conformal field theories in $d=4$\\ with a helical twist}
  \vskip 1.5cm
  \vskip 1.5cm
      Aristomenis Donos$^1$, Jerome P. Gauntlett$^2$ and Christiana Pantelidou$^3$\\
   \vskip .6cm
    \vskip .6cm
      \begin{small}
      \textit{$^1$Centre for Particle Theory, Department of Mathematical Sciences, \\
Durham University,  Durham DH1 3LE, U.K.}
        \end{small}\\
      \vskip .6cm
      \begin{small}
      \textit{$^2$Blackett Laboratory, 
        Imperial College\\ London, SW7 2AZ, U.K.}
        \end{small}\\
          \vskip .6cm
      \begin{small}
      \textit{$^3$Departament de F\'isica Fonamental (FFN) and Institut de Ci\`encies del Cosmos (ICC)\\ Universitat de Barcelona (UB), Mart\`i i Franqu\`es 1, E-08028 Barcelona Spain}
        \end{small}\\*[.6cm]
         
\end{center}

\vfill

\begin{center}
\textbf{Abstract}
\end{center}

\begin{quote}
Within the context of holography we study the general class of $d=4$ conformal field theories (CFTs) after applying
a universal helical deformation.
At finite temperature we construct the associated black hole solutions of Einstein gravity, numerically,
by exploiting a Bianchi $VII_0$ ansatz for the bulk $D=5$ metric. At $T=0$ we show that they flow in the IR to
exactly the same CFT. The deformation gives rise to a finite, non-zero DC
thermal conductivity along the axis of the helix, which we determine analytically in terms of black hole horizon data. 
We also calculate the AC thermal conductivity along this axis and
show that it exhibits Drude-like peaks.
\end{quote}

\vfill

\end{titlepage}
\setcounter{equation}{0}

\section{Introduction}
There has been significant recent interest in constructing black hole solutions that are holographically
dual to conformal field theories (CFTs) deformed by operators which break translation
invariance. One motivation for these studies is that the UV deformation provides a mechanism by which momentum can be dissipated
in the conformal field theory giving rise to more realistic transport behaviour without delta functions \cite{Hartnoll:2012rj,Horowitz:2012ky,Horowitz:2012gs,Donos:2012js,Chesler:2013qla,Ling:2013nxa,Donos:2013eha,Andrade:2013gsa,Balasubramanian:2013yqa,Donos:2014uba,Gouteraux:2014hca,Donos:2014cya,Donos:2014oha,Donos:2014yya,Kim:2014bza,Davison:2014lua}\footnote{Another approach to study momentum dissipation
is by using massive gravity e.g. \cite{Vegh:2013sk,Blake:2013owa,Davison:2013txa,Amoretti:2014mma}.}. 
A second and interrelated motivation is that it provides a framework for seeking new ground states, both metallic and insulating, as well
as transitions between them \cite{Donos:2012js,Donos:2013eha,Donos:2014uba,Gouteraux:2014hca,Baggioli:2014roa}.

In this paper we analyse a specific helical deformation of $d=4$ conformal field theories which is appealing both because it is universal
and because it is possible, at a technical level, to analyse it in some detail. 
The deformation consists of
a helical source for the energy-momentum tensor of the CFT which breaks the spatial Euclidean symmetry down to
a Bianchi VII$_0$ subgroup. This deformation is equivalent to considering the CFT not on four-dimensional Minkowski
spacetime, but on the spacetime with line element
\begin{align}\label{bdymet}
ds^{2}&=-dt^{2}+\omega_{1}^{2}+e^{2\alpha_0}\,\omega_{2}^{2}+e^{-2\alpha_0}\,\omega_{3}^{2}\,.
\end{align}
Here $\omega_i$ are the left-invariant one-forms associated with 
the Bianchi VII$_0$ algebra,
\begin{equation}\label{liofs}
\omega_1=dx_1, \quad \omega_2=\cos kx_1 dx_2-\sin k x_1 dx_3,\quad \omega_3=\sin kx_1 dx_2+\cos k x_1 dx_3\,,
\end{equation}
with constant wave-number $k$,
and the constant $\alpha_0$ parametrises the strength of the helical deformation.

The black holes that are dual to these helical deformations at finite temperature are all
solutions for the $D=5$ Einstein-Hilbert action with negative cosmological constant, and hence are relevant for
the entire class of $d=4$ conformal field theories with an $AdS_5$ dual. As we will see, the ansatz for the 
$D=5$ metric is static and maintains the Bianchi 
VII$_0$ symmetry and hence constructing explicit black hole
solutions just requires solving ordinary differential equations (ODEs) in the radial variable\footnote{Bianchi VII$_0$ symmetry has arisen in various holographic constructions 
\cite{Nakamura:2009tf,Donos:2011ff,Iizuka:2012iv,Donos:2012gg,Donos:2012wi,Iizuka:2012pn,Donos:2013woa} including constructions with momentum dissipation \cite{Donos:2012js,Donos:2014oha}.}.

Classically, the UV deformation parameter $\alpha_0$ in \eqref{bdymet} is a dimensionless
number, but due to the conformal anomaly, a nontrivial dynamical scale is introduced.
For a fixed dynamical scale, the system depends on the value of
$\alpha_0$ and on the dimensionless ratio $k/T$. When $T=0$, we can study the effect of the $\alpha_0$ deformation by considering a perturbative analysis about
the $AdS_5$ vacuum. We find that the solution approaches exactly the same $AdS_5$ vacuum solution in the IR, with a simple renormalisation of length scales.
This is very similar to what is seen for the deformation of $d=3$ CFTs by a periodic chemical potential which averages to zero 
over a period \cite{Chesler:2013qla}, and is also reminiscent of some ground states of particular 
$s$-wave \cite{Horowitz:2009ij} and $p$-wave \cite{Basu:2010fa,Donos:2013woa} superconductors.

To go beyond this perturbative analysis, and also to consider $k/T\ne0$, we construct fully back-reacted black hole solutions
using a numerical shooting method and study their properties. 
For all values of $\alpha_0$ that we have considered, we show that as $T/k\to 0$ the black holes
approach $T=0$ solutions which 
interpolate between $AdS_5$ in the UV and the same $AdS_5$ in the IR, just as in the perturbative analysis. 
In particular, we find that for the ranges of parameters that we have considered, the deformations do not lead to any new ground states. 

Following the construction of the black hole solutions we calculate the thermal AC conductivity as a function of frequency, 
$\kappa(\omega)$, by calculating the two-point function
for the momentum operator $T^{tx_1}$. In fact the operators $T^{tx_1}$ and
$T^{\omega_2\omega_3}$ mix and we calculate the full two by two matrix of AC conductivities, including contact terms.
This calculation requires a careful treatment of gauge-transformations
and we employ, and also further
develop, the method used in \cite{Donos:2013eha,Donos:2014yya}. We observe Drude peaks in $\kappa(\omega)$, with an associated nonvanishing DC conductivity at finite $T/k$.

We also derive an analytic expression for the associated DC thermal conductivity in terms of the black hole horizon data, using the technique of 
\cite{Donos:2014uba,Donos:2014cya}. In addition, analytic expressions for the other components of the matrix of DC conductivities are obtained in terms of UV data of the background black hole solutions. The DC calculation also
leads to concise expressions for the static susceptibilities, the Green's functions at zero frequency, which agrees with
the limit of the AC results.

\section{Black hole solutions}\label{sec:BH}
We consider the five-dimensional Einstein-Hilbert action given by 
\begin{equation} \label{eq:action}
S=\int d^5x \sqrt{-g}(R+12)\,,
\end{equation}
where we have set $16\pi G=1$ and fixed the cosmological constant to be $\Lambda=-6$ for convenience. 
The equations of motion are simply given by 
\begin{equation}\label{eeqs}
R_{mn}=-4 g_{mn}\,,
\end{equation}
and admit a unique $AdS_5$ vacuum solution, with unit radius, which is dual to a $d=4$ CFT.

The metric ansatz for the black hole solutions that we shall consider is given by 
\begin{align}\label{eq:ansatz}
ds^{2}&=-g\,f^{2}\,dt^{2}+g^{-1}{dr^{2}}+h^{2}\,\omega_{1}^{2}+r^{2}\,\left(e^{2\alpha}\,\omega_{2}^{2}+e^{-2\alpha}\,\omega_{3}^{2}\right)\,,
\end{align}
where $g, f , h,\alpha$ are all functions of the radial coordinate, $r$, only and $\omega_i$ are the left-invariant one-forms associated with 
the Bianchi VII$_0$ algebra given in \eqref{liofs}.
Clearly this ansatz is static with a Bianchi VII$_0$ symmetry. 
After substituting the ansatz into \eqref{eeqs} we obtain the following system of ODEs:
\begin{align}\label{els}
f'+\frac{f \left(-2 r h^2 \left(g \alpha'^2+2\right)+r \left(2 k^2 \sinh ^22 \alpha-g h'^2\right)+\left(g+4 r^2\right) h h'\right)}{g h \left(r h'+2 h\right)}&=0\,,\nn
g'+\frac{2 \left[h^2 \left(r^2 g \alpha'^2+g-2 r^2\right)+r^2 \left(g h'^2-k^2 \sinh ^22 \alpha\right)+r \left(g-4 r^2\right) h h'\right]}{r h \left(r h'+2 h\right)}&=0\,,\nn
h''+\frac{4 r h'}{g}-\frac{4 h}{g}+\frac{h'}{r}-\frac{h'^2}{h}+\frac{2 k^2 \sinh ^22 \alpha}{g h}&=0\,,\nn
\alpha''+\frac{4 r \alpha'}{g}+\frac{\alpha'}{r}-\frac{k^2 \sinh 4 \alpha}{g h^2}&=0\,.
\end{align}
For future reference, note
that the AdS-Schwarzschild black hole solution, describing
the CFT at finite temperature $T$ with no deformation, has $g=r^2-\frac{r_+^4}{r^2}, f=1, h=r$ 
and $\alpha=0$, $k=constant$\footnote{Alternatively,
one can set $k=0$ and $\alpha=constant$, which can then be scaled to zero.}, 
 with $T=r_+/\pi$.

Observe that the ansatz, and hence the equations of motion, preserves the parity transformation 
$(x_1,k)\to -(x_1,k)$, and is also invariant under the following three scaling symmetries:
\begin{align}
\label{eq:symmetries}
&r \to \lambda r\,, \quad (t,x_2,x_3) \to \lambda^{-1}(t,x_2,x_3)\,, \quad g\to \lambda^2 g\,; \nonumber\\
&x_1 \to \lambda^{-1} x_1\,, \quad h \to \lambda h\,,  \quad  k \to \lambda k\,;\nonumber\\
&t \to \lambda t\,, \quad f \to \lambda^{-1}f\,; \end{align}
where $\lambda$ is a constant.

\subsection{UV and IR expansions}
We now discuss the boundary conditions that we will impose on \eqref{els}.
In the UV, as $r\to\infty$, we demand that we have the asymptotic behaviour given by
\begin{align} 
f=& f_{0}\Big(1+ \frac{k^2}{12 r^2}(1-\cosh 4\adef)-\frac{c_h}{r^4}+\frac{k^4}{96 r^4}(3+4\cosh 4 \adef-7\cosh 8 \adef)\notag\\
&\qquad\qquad\qquad\qquad\qquad\qquad\qquad-\frac{k^4 \log r}{6 r^4}(\cosh 4 \adef-\cosh 8 \adef)+\cdots \Big),\notag\\
g=& r^{2}\left(1-\frac{k^2}{6 r^2}(1-\cosh 4 \adef)-\frac{M}{r^4}+\frac{k^4 \log r}{3 r^4}(\cosh 4 \adef-\cosh 8 \adef)+\cdots\notag\right),\notag\\
h=& r\,\left(1-\frac{k^2}{4 r^2} (1-\cosh 4\adef) +\frac{c_h}{r^4}+\frac{k^4 \log r}{6 r^4}(\cosh 4\adef-\cosh 8 \adef)+\cdots \right),\notag\\
\alpha=&\adef-\frac{k^2}{4 r^2}\sinh 4 \adef+\frac{c_\alpha}{r^4}-\frac{k^4 \log{r}}{12 r^4}( \sinh{4 \adef}-2 \sinh{8 \adef}) +\cdots.
\label{eq:expUV}
\end{align}
The most important thing to notice is that this implies that the metric is approaching $AdS_5$ with a helical deformation, with pitch $2\pi/k$, 
that is parametrised by $\alpha_0$ as in \eqref{bdymet}. 
The expansion \eqref{eq:expUV} is, in fact, specified in terms of six parameters $M,f_0,c_h, \adef,c_\alpha$ and $k$.  
The third scaling symmetry in \eqref{eq:symmetries} allow us to set $f_0=1$ and we will do so later on. 
Note that the second scaling symmetry in \eqref{eq:symmetries} is not preserved by the UV ansatz. However, when combined with the first
we deduce that under $r\to \lambda r$ and rescaling the field theory coordinates by $\lambda^{-1}$ 
the ansatz is preserved by the following scaling symmetries of the UV parameters:
\begin{align}\label{anomscal}
f_0& \to f_0\,,\qquad
\adef \to \adef \,,\qquad
k \to \lambda k\,,\notag\\
M & \to  \lambda^4 M+\frac{ (\lambda k)^4}{3}(\cosh 4 \adef-\cosh 8 \adef)\log \lambda\,,\notag\\
c_h & \to \lambda^4 c_h-\frac{(\lambda k)^4 }{6}(\cosh 4\adef-\cosh 8 \adef)\log \lambda\,,\notag\\
c_\alpha & \to \lambda^4 c_\alpha+\frac{(\lambda k)^4 }{12}( \sinh{4 \adef}-2 \sinh{8 \adef})
\log{\lambda}\,.
\end{align} 
The log terms are associated with an anomalous scaling of physical quantities due to the conformal anomaly.

In the IR, we assume that we have a regular black hole Killing horizon located at $r=r_+$. We 
thus demand that as $r\to r_+$ we can develop the expansion:
\begin{align}\label{bhas}
&g=g_+(r-r_+)- \frac{4h_+^2+k^2(1-\cosh 4\alpha_+)}{2 h_+ ^2}(r-r_+)^2+\cdots\,,\notag\\
&f=f_++0(r-r_+)+\cdots\,,\notag\\
&h=h_+  + \frac{4h_+^2+k^2(1-\cosh 4\alpha_+)}{4 h_+ r_+}(r-r_+)+\cdots\,,\notag\\
&\alpha=\alpha_++\frac{k^2\sinh 4\alpha_+}{4h_+^2r_+}(r-r_+)+\cdots\,.
\end{align} 
This expansion is specified in terms of four parameters $r_+, f_+,h_+$ and $\alpha_+$, with $g_+$ fixed to be $g_+=4r_+$. 

The equations of motion \eqref{els} consist of two first-order equations for $g,f$ and two second-order equations for $h,\alpha$ and hence a solution is specified by six constants of integration. On the other hand, we have ten parameters in the boundary conditions minus two
for the remaining scaling symmetries \eqref{eq:symmetries}.
We thus expect to find a two parameter family of solutions parametrised by the deformation parameter $\alpha_0$ and $k/T$, both of which are dimensionless. Note that the presence of the conformal anomaly introduces an additional dynamical energy scale into the system which we will
hold fixed to be unity throughout our analysis.

\subsection{Thermodynamics}\label{sec:Thermodynamics}
To analyse the thermodynamics of the black  hole solutions we need to calculate the on-shell Euclidean action. We analytically  continue the time coordinate by setting $t=-i\tau$ . Near $r=r_+$, the Euclidean solution takes the approximate form
\begin{align}
\label{eq:ansatz2}
&ds_E^{2}\approx g_+ f_+^2(r-r_+)d\tau ^2+\frac{dr^2}{g_+(r-r_+)}+h_+^2 dx_1^2+r_+^2(e^{2 \alpha_+}\omega_2^2+e^{-2 \alpha_+}\omega_3^2)\,.
\end{align}
The regularity of the solution at $r=r_+$ is ensured by demanding that
 $\tau$ is periodic with period $\Delta \tau=4 \pi /(g_+f_+)$, corresponding to temperature $T=(f_0 \Delta \tau)^{-1}$. We can also read off the area of the event horizon  and since we are working in units  with $ 16 \pi G=1$, we deduce that the entropy density is given by 
\begin{equation}
s=4 \pi r_+^2 h_+\,.
\end{equation}
Following \cite{Henningson:1998gx,Balasubramanian:1999re,Emparan:1999pm,Taylor:2000xw,deHaro:2000xn} 
we will consider the total Euclidean action, $I_{Tot}$, defined as
\begin{equation}
I_{Tot}=I+I_{ct}+I_{Log}\,,
\end{equation}
where $I=-iS$ and $I_{ct}$  is given by the following integral on the boundary $r\to \infty$:
\begin{equation}\label{ctermp}
I_{ct}=\int d\tau d^3 x \sqrt{-\gamma}(-2 K+6+\frac{1}{2}R)\,.
\end{equation}
Here $K$
is the trace of the extrinsic curvature of the boundary, $\gamma_{\mu\nu}$ is the induced boundary metric and $R$ is the associated Ricci scalar.
$I_{Log}$ is also needed for regularising the action and is given by
\begin{equation}\label{ctermlog}
I_{Log}=\int d\tau d^3 x \sqrt{-\gamma}\frac{\log r}{8} (-\frac{2}{3}R^2 + 2 R_{\mu\nu} R^{\mu\nu})\,,
\end{equation}
where $R_{\mu\nu}$ is the Ricci tensor associated with $\gamma_{\mu\nu}$.
For our ansatz, with the induced boundary line element associated with $\gamma_{\mu\nu}$ given by
$r^2$ times the metric in \eqref{bdymet},
\begin{align}\label{bdymet2}
ds^2_{\infty}=r^2\left(-dt^2+dx_1^2+
e^{2\alpha_0}\,\omega_{2}^{2}+e^{-2\alpha_0}\,\omega_{3}^{2}\right)\,,
\end{align}
we have
\begin{align}\label{ctermpex}
I_{ct}=&Vol_3 \Delta \tau \lim_{r \to \infty} r^2 h f g^{1/2}[6-2g^{1/2} (\frac{2}{r} +\frac{f'}{f}+ \frac{h'}{h})- g^{-1/2} g'
-\frac{k^2\sinh^22\alpha}{ h^2}]\,,\nonumber \\
I_{Log}=&Vol_3 \Delta \tau \lim_{r \to \infty} \frac{r^2  f g^{1/2}k^4}{3h^3}   \log r[\cosh8\alpha-\cosh4\alpha]\,,
  \end{align}
where $Vol_3=\int dx_1 dx_2 dx_3$. We next point out two equivalent ways to write the bulk part of the Euclidean action on-shell:
\begin{align}\label{act2waysp}
I_{OS}=& Vol_3 \Delta\tau \int_{r_+}^\infty dr \left(2 r g h f\right)'\,,\nonumber \\
          =& Vol_3 \Delta\tau \int_{r_+}^\infty dr \left(r^{2}hfg^{\prime}+2r^{2}hgf^{\prime}\right)'\,.
        \end{align}
Notice that the first expression only receives contributions from the boundary at $r \to \infty$ since $g(r_+)=0$, while the second expression also receives contributions from $r=r_+$. 
We next define the free energy $W=T[I_{Tot}]_{OS}\equiv w Vol_3$.
Using the UV and IR expansions we obtain the following expression for the free energy density:
\begin{align}
\label{eq:OSactionp}
w&= -M+\frac{k^4}{12}\sinh^42\alpha_0\, ,\nonumber \\
          &= 3M +8 c_h -s T+\frac{k^4}{24}\sinh^2 2\alpha_0(35+61\cosh 4\alpha_0)\,,
          \end{align}
and hence the Smarr-type formula:
\begin{align}\label{smarrp}
4M+8 c_h -s T+\frac{k^4}{2}\sinh^2 2\alpha_0\left(3+5\cosh{4 \adef}\right)=0\,.
\end{align}
Observe that under the scaling \eqref{anomscal}, the log terms drop out of \eqref{smarrp}.

We now compute the expectation value of the boundary stress-energy tensor following \cite{Balasubramanian:1999re}. 
The relevant terms are given by\footnote{Note that we will write
$T^{\mu\nu}=r^6\tilde{T}^{\mu\nu}$ with $T^{\mu\nu}$ then independent of $r$.}
\begin{equation}\label{stressy}
\langle \tilde{T}^{\mu\nu}\rangle\equiv\lim_{r\to\infty} [-2 K^{\mu\nu}+2(K-3)\gamma^{ \mu\nu}+R^{\mu \nu}-\frac{1}{2}\gamma^{ \mu\nu}R-\log r(\frac{K_1^{\mu \nu}}{4}-\frac{K_{2}^{\mu\nu}}{12})+\cdots]\,,
\end{equation}
where 
\begin{align}
K_{1}^{\mu \nu}=&\frac{2}{\sqrt{-\gamma}}\frac{\delta}{\delta \gamma_{\mu \nu}}[\sqrt{- \gamma} R_{\rho\sigma} R^{\rho\sigma}]\,,\qquad
K_{2}^{\mu \nu}=\frac{2}{\sqrt{-\gamma}}\frac{\delta}{\delta \gamma_{\mu \nu}}[\sqrt{- \gamma} R^2 ]\,.
\end{align}
and explicit formulas can be found in \cite{Ford:1997hb}. 
Using the asymptotic expansion \eqref{eq:expUV}, one obtains the boundary stress-energy tensor, which we present in Appendix \ref{ap:StressTensor}. It is straightforward to explicitly show that the stress tensor is conserved, $\nabla_\mu \tilde T^{\mu\nu}=0$, where the covariant derivative is respect to the boundary metric \eqref{bdymet2}, as expected. We can also calculate the trace 
of the energy-momentum tensor with respect to the boundary metric to get the conformal anomaly\footnote{Note that in the present setup
the boundary metric satisfies $\Box R=0$ and there is no ambiguity in the trace of the stress tensor.} 
\begin{align}\label{tracebdy}
\tilde{T}^{\mu}{}_{\mu}=&-{2}\left( \frac{1}{24}R^2-\frac{1}{8} R_{ij} R^{i j}\right)\,,\nonumber\\
=&\frac{1}{r^{4}}\frac{k^4}{3}\left(\cosh(8 \adef)-\cosh(4\adef)\right)\,.
\end{align}
We also note that with the stress tensor in hand, as further discussed in Appendix \ref{ap:StressTensor},
we can use the results of \cite{Donos:2013cka} to immediately recover the 
Smarr formula \eqref{smarrp} and also the first law
\begin{align}\label{fstlaw}
\delta w=&-s\delta T+\left(8 c_h+2k^4\sinh^2 2\alpha_0(1+ 2 \cosh 4\alpha_0)\right)\frac{\delta k}{k}\nn
&-\frac{1}{4}\left( 64 c_\alpha + 3k^4(2\sinh 4\alpha_0-3\sinh 8\alpha_0)\right)\delta \alpha_0\,.
\end{align}

\subsection{Perturbative helical deformation about $AdS_5$}\label{pertexp}
Before constructing the back-reacted black hole solutions, it is illuminating to investigate, within perturbation theory, the impact of the helical deformation
about $AdS_5$ space-time (at $T=0$).
Specifically, we focus on the small $\alpha$ deformation given by
\begin{align}
g&=r^2+\epsilon^2 \delta g\,,\qquad
f=1+\epsilon^2 \delta f\,,\qquad
h=r+\epsilon^2 \delta h\,,\qquad
\alpha=0+\epsilon \delta \alpha\,.
\end{align} 
At first order in $\epsilon$, the Einstein equations \eqref{els} imply
\begin{equation}
\label{eq:per1order}
\delta \alpha ''+\frac{5}{r} \delta \alpha' -\frac{4 k^2}{r^4}\delta \alpha=0\,,
\end{equation}
while at second order we obtain
\begin{align}
\label{eq:per2order}
&\delta g'+\frac{2}{r} g-\frac{8 k^2}{3r} \delta \alpha^2+\frac{2 r^3}{3} \delta \alpha'^2=0\,,\nonumber\\
&\delta f'+\frac{1}{r^2} (r\delta h'-\delta h)+\frac{8 k^2}{3r^3}\delta \alpha^2-\frac{2 r}{3} \delta \alpha'^2=0\,,\nonumber\\
&\delta h''+\frac{3}{r} \delta h'-\frac{3}{r^2}\delta h+\frac{8 k^2}{r^3}\delta \alpha^2=0\,.
\end{align}
Equation \eqref{eq:per1order} can be solved analytically in terms of Bessel functions. Demanding that it is regular at the Poincar\'e horizon at $r=0$ and
that it approaches $\adef$ as $r\to \infty$ implies that 
\begin{equation}\label{blipp}
\delta \alpha=\frac{2k^2\alpha_0}{r^2}K_2(2k/r)\,.
\end{equation}
Given this, one can then solve \eqref{eq:per2order} in terms of Meijer G-functions, again subject to regularity at the horizon and suitable asymptotic fall-off.
Given the analytic solutions obtained, we find that in the far IR the metric becomes approximately
\begin{equation}
ds^2(IR)\approx-r^2dt^2+\frac{dr^2}{r^2}+r^2(1+\frac{8\adef^2}{5})dx_1^2+r^2(dx_2^2+dx_3^2)\,.
\end{equation} 
Thus, we see that at $T=0$, the effect of a small $\alpha_0$ deformation does not change the IR behaviour away from the unit radius $AdS_5$, apart from a renormalisation of length scales in the $x_1$ direction given by
\begin{equation}\label{renp}
\bar\lambda\equiv
\sqrt{\frac{g_{x_1x_1}(r\to0)}{g_{x_1x_1}(r\to\infty)}}=1+\frac{4\adef^2}{5}\,.
\end{equation}
In section \ref{dws} we will explicitly construct the full non-linear $T=0$ solutions which interpolate between $AdS_5$ in the UV and $AdS_5$ in the IR, that agree with both the perturbative analysis just discussed, as well as the $T\to 0$ limit of the black holes that we now discuss.

\subsection{Numerical construction of black hole solutions} \label{sec:BHs}
We construct the black hole solutions numerically. We solve the system of ODEs \eqref{els} with the
asymptotic behaviour given by \eqref{eq:expUV} and \eqref{bhas}
using a shooting method, for fixed values of $\alpha_0$ and $k$ and then cool them down to low temperatures. Recall that for a given dynamical
scale, the parameters specifying the UV data are $\alpha_0$ and the dimensionless ratio $T/k$. In practice we
set $k=1$.

\begin{figure}
\centering
{\includegraphics[width=7cm]{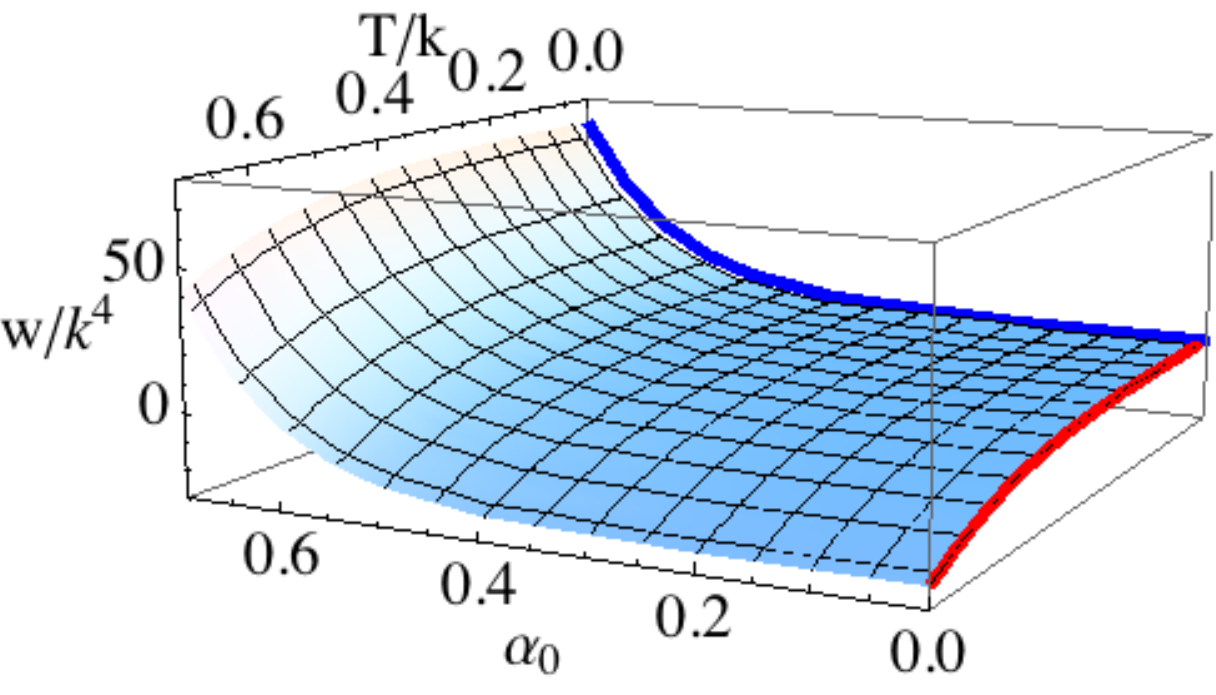}}\quad
{\includegraphics[width=7cm]{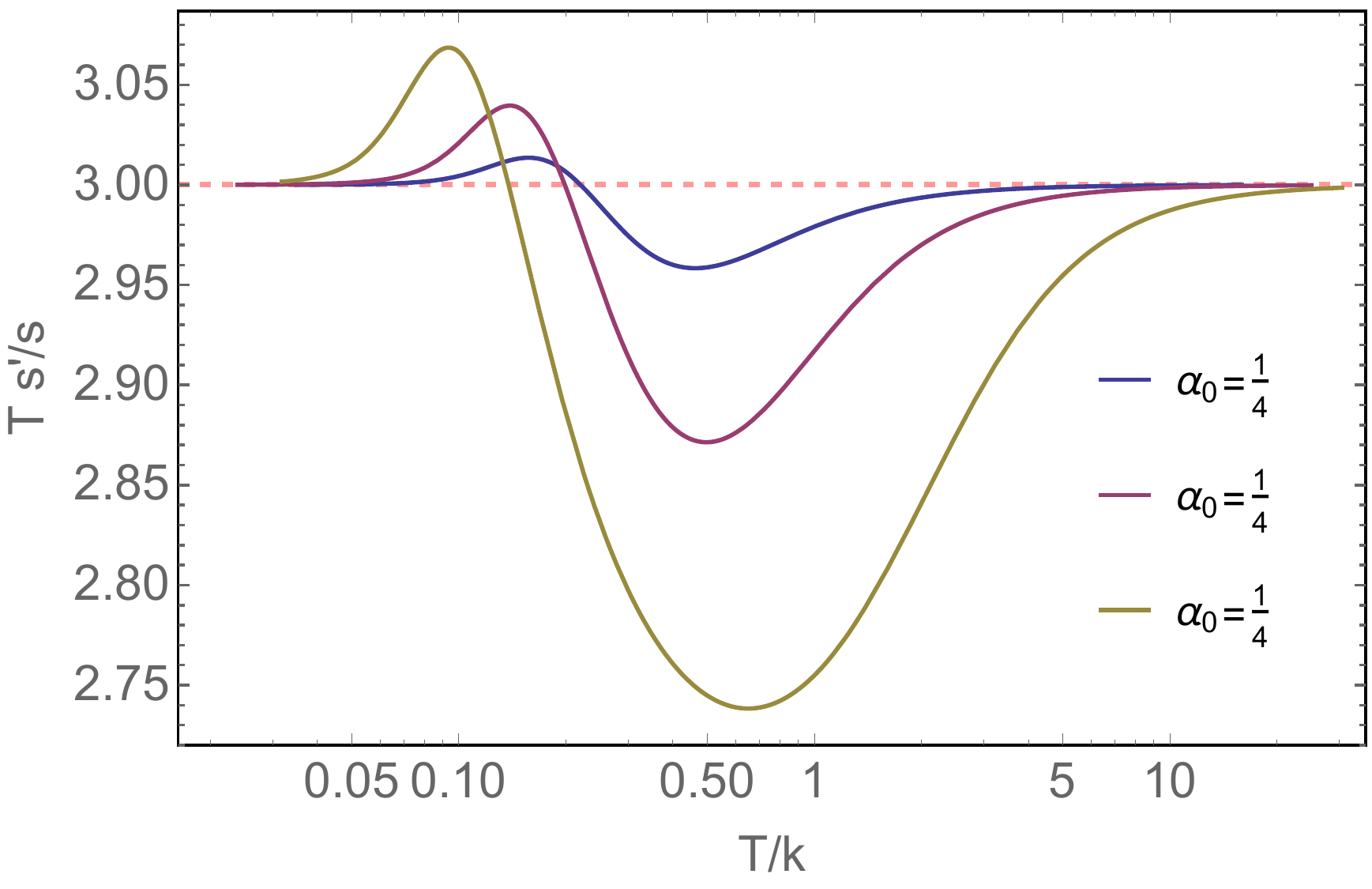}}
\caption{
The left panel plots the free energy, $w$, of the black holes constructed versus the deformation parameter, $\alpha_0$, and temperature, $T$, and we have divided  by suitable powers of $k$ to plot dimensionless quantities. 
The blue line is the free energy of the $T=0$ solutions, which were constructed independently, 
and the red line is the free energy of the AdS-Schwarzschild black holes.
The right panel is a plot of the entropy exponent, given by $T s'/s$,  against $T/k$ for the helical deformed black holes with $\alpha_0=1/4$ (blue),
$\alpha_0=1/2$ (purple) and $\alpha_0=1$ (olive). For low temperatures the exponent approaches three, associated with the reappearance of $AdS_5$ in the far IR.}\label{fig:DWs}
\end{figure}

In the left panel of figure \ref{fig:DWs} we display the free energy of the black hole solutions as a function of the deformation
parameter $\alpha_0$ and the dimensionless temperature ratio, $T/k$. As $\alpha_0\to 0$, with the deformation being switched off, the solutions
smoothly approach the AdS-Schwarzschild solution, as expected. For example, 
on figure \ref{fig:DWs} the red line denotes the free-energy of the AdS-Schwarzschild solution (with $T/k\to T$).

We can also examine the behaviour of the solutions as the temperature is lowered to zero, $T/k\to 0$. An examination of the solutions shows that the entropy is going to zero with the power-law behaviour $s\sim T^3$. This behaviour is clearly seen in the right panel of figure \ref{fig:DWs} where we have plotted $T s'/s$ for three different deformation parameters, 
$\alpha_0=1/4,1/2$ and $1$.
This behaviour suggests that all of the black holes are approaching $AdS_5$ in the far IR at $T=0$. 
This conclusion is supported by an analysis of the behaviour of the functions entering the metric. 
It is further supported by an explicit construction of $T=0$ solutions that interpolate
between the same $AdS_5$ in the UV and the IR, which we discuss in the next subsection.
For example,
the blue line in the left panel of figure \ref{fig:DWs} shows that the free energy of the black holes agrees with that of
the $T=0$ solutions.

\subsection{Non-linear helical deformation about $AdS_5$ at $T=0$}\label{dws}
We now construct $T=0$ solutions\footnote{Note that we do not call these solutions ``domain wall" solutions because the same
$AdS_5$ vacua is present at each end of the RG flow.} that interpolate between $AdS_5$ in the UV and 
in the IR. We find that their properties are precisely consistent with the $T\to 0$ limit of the black hole
solutions of the last subsection.

The UV expansion is the same as we had for the black holes given in \eqref{eq:expUV}. 
In the IR, as $r\to 0$, building on \eqref{blipp}, we use the following double expansion
\begin{align}\label{eq:expIRDW}
&g=r^2 + \frac{k^3\bar \alpha_+^2 }{r}e^{-4 k/\bar h_+ r}(1+\frac{5\bar h_+}{8k}r+{\cal O}(r^2))+\cdots\,,\notag\\
&f=\bar f_+-\frac{k^3\bar \alpha_+^2 \bar f_+}{2 r^3}e^{-4 k/\bar h_+ r}(1+\frac{5\bar h_+}{8k}r+{\cal O}(r^2))+\cdots\,,\notag\\
&h=\bar h_+ r-\frac{k^3\bar\alpha_+^2 \bar h_+ }{2 r^2}e^{-4 k/\bar h_+ r}(1+\frac{21\bar h_+}{8k}r+{\cal O}(r^2))+\cdots\,,\notag\\
&\alpha=\frac{\bar \alpha_+ 2k^2}{ \sqrt{\pi \bar h_+}r^{2}} K_2\left (\frac{2 k}{\bar h_+ r}\right)+\cdots\,,
\end{align} 
where the neglected terms are ${\cal O}(e^{-6 k/\bar h_+ r}$).
In the far IR the metric approaches the $AdS_5$ vacuum solution with the flow governed by $k$-dependent relevant modes, in the same spirit as in \cite{Chesler:2013qla,Donos:2013woa}. This IR expansion is specified by three dimensionless constants $\bar \alpha_+,\bar f_+,\bar h_+$ and we observe, in particular, that
$\bar f_+$ and $\bar h_+$ allow for a nontrivial renormalisation of length scales between the UV and the IR.
By a simple counting argument, we expect to find a one parameter family of solutions parametrised by the deformation parameter $\alpha_0$ (for a fixed dynamical scale).

\begin{figure}[h]
\centering
{\includegraphics[width=7cm]{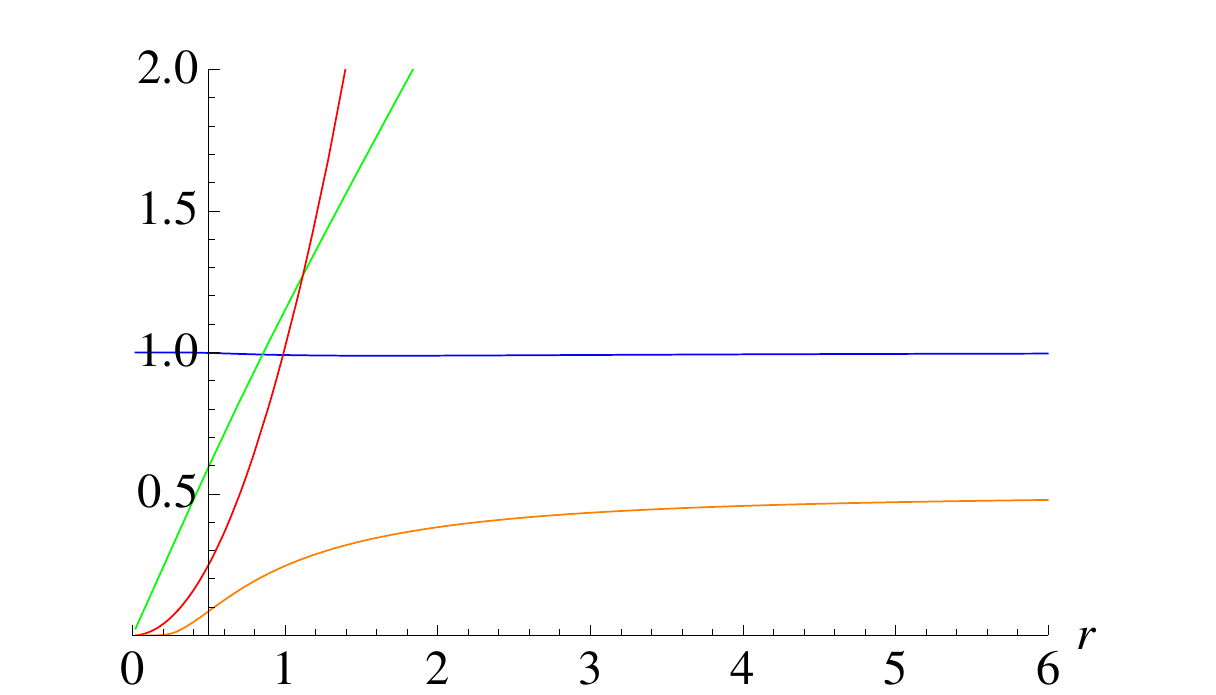}}\qquad
{\includegraphics[width=7cm]{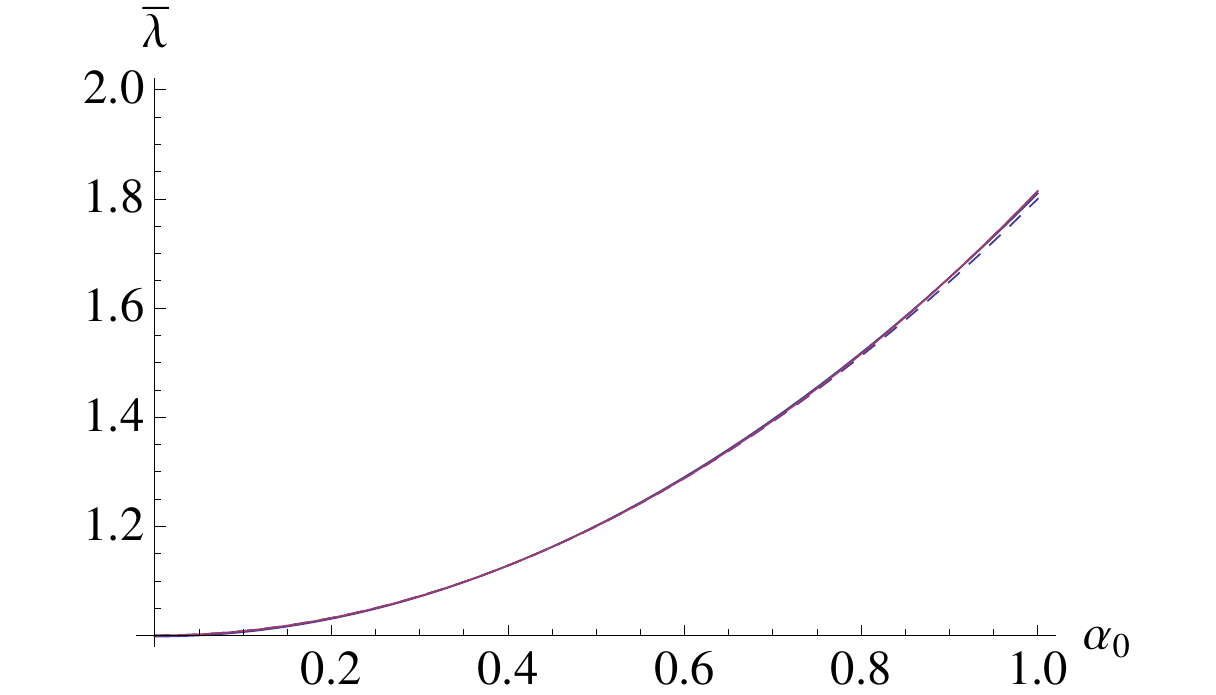}} 
\caption{Plots for the $T=0$ helical deformation solutions 
about $AdS_5$. The left panel plots the behaviour of the metric functions
$g$ (red), $f$ (blue), $h$ (green) and $\alpha$ (orange) 
against radius for the case of $\alpha_0=1/2$. We see that the solution smoothly
approaches $AdS_5$ in the far IR. The right panel plots the parameter $\bar\lambda$, defined in \eqref{renp}, which captures the renormalisation of length scales, for various $\alpha_0$. 
The almost coincident dashed line is the behaviour obtained using the perturbative expansion in section \ref{pertexp}.}
\label{fig:TZ}
\end{figure}

We proceed by solving the equations of motion subject to the above boundary conditions, again using a shooting method.
The behaviour of the functions is summarised in the left panel of figure \ref{fig:TZ} for $\alpha_0=1/2$
and we see that they smoothly interpolate between the same $AdS_5$ in the UV and IR.
Similar behaviour is also seen for other values of $\alpha_0\ne0$. To see that these solutions are indeed the $T\to 0$ limit of
the black holes constructed in the previous subsection, we can compare the expectation values in the UV data of the domain wall solutions with those of the black hole solutions 
and we find precise agreement. For example, in the left panel of 
figure \ref{fig:DWs} we display the free energy density.

It is also interesting to note that the expansion \eqref{eq:expIRDW}, combined with the analytic AdS-Schwarzschild black hole
(given just below \eqref{els}), allows us to obtain the low-temperature scaling behaviour of the finite temperature black holes.
At low temperatures the radius of the black hole horizon will be related to the temperature via $r_+\sim\pi T/\bar f_+$. Thus,
the low-temperature scaling of the entropy density, for example, is given by a double expansion of the form
\begin{equation}\label{scess}
s=4\pi r_+^2 h(r_+)\sim\frac{4 \bar h_+ \pi^4 T^3}{\bar f_+^{3}}\left(1 -\frac{ \bar \alpha_+^{2} \bar f_+^{3}}{2 \pi^3 T^3} e^{-4 \bar f_+ k/\bar h_+ \pi T}+\dots \right)+\dots\,,
\end{equation}
where the parameters on the right-hand side are as in \eqref{bhas}.

In section \ref{pertexp} we have seen in the perturbative analysis for small $\alpha_0$ that
there is a renormalisation of the length scale in going from the UV to the IR. 
By constructing the $T=0$ solutions for various $\alpha_0$ we can plot the dependence of the renormalisation parameter
$\bar\lambda$, defined in \eqref{renp}, as we vary $\alpha_0$, as shown in figure \ref{fig:TZ}.

\section{Two-point functions of the stress tensor}

In this section we calculate two-point functions of the $T^{t x_1}$ and $T^{\omega_2\omega_3}$
components of the stress tensor, which mix in the $\alpha$-deformed helical backgrounds, 
at finite frequency, $\omega$, and zero spatial momentum. In particular, from the $T^{t x_1}T^{t x_1}$ correlator we can 
obtain the AC thermal conductivity, $\kappa(\omega)$. We will use and further develop the method discussed
in \cite{Donos:2013eha,Donos:2014yya}, building on \cite{Son:2002sd,Kaminski:2009dh}.

\subsection{The perturbation}\label{thepert}
We consider the following time-dependent perturbations, $ds^2\to ds^2+\delta (ds^2)$,
around the background solutions that we constructed in section \ref{sec:BH} with
\begin{equation}
\label{eq:deltag}
\delta (ds^2)=2\delta g_{t x_1}(t,r) dt dx_1+2\,\delta g_{2 3}(t,r)\omega_2 \omega_3\,.
\end{equation}
It will be convenient to Fourier decompose our perturbations as 
\begin{equation}\label{fourts}
\delta g_{tx_1}(t,r)=\int\frac{d\omega}{2\pi} e^{-i \omega t}h_{tx_1}(\omega,r)\,,\qquad
\delta g_{23}(t,r)=\int\frac{d\omega}{2\pi} e^{-i \omega t}h_{23}(\omega,r)\,,
\end{equation}
and the reality of the perturbation implies that 
\begin{align}\label{reality}
\bar h_{tx_1}(\omega,r)=h_{tx_1}(-\omega,r),\qquad \bar h_{23}(\omega,r)=h_{23}(-\omega,r)\,.
\end{align}
It is straightforward to show that this perturbation gives rise to two second-order equations
\begin{align}\label{sceqs}
&h_{2\,3}''-\frac{1}{r^2 f^2 g^2 h^2}\Big(- 2i\omega h_{t x_1} k r^4\sinh 2 \alpha +h_{2\,3}'f^2 g h^2 (-4 r^3+3 r g)\nonumber\\
&+ h_{2\,3}[-r^2 \omega^2  h^2+2 f^2 gk^2 r^2(1+\cosh{4\alpha})+4 f^2 g h^2(2 r^2-g+gr^2 \alpha'^2)]\Big)=0\,,\nonumber\\
&h_{t\,x}''-\frac{1}{r^2 g h^2 (2h+rh')}\Big( h_{t\,x}4 r(4 r h^3+k^2 r^2h' \sinh^2 2\alpha+2r ghh'^2-4 r^2h^2h'+h^2h'g-r^2 h^2h'g\alpha'^2)\nonumber\\
&+h_{t\,x}' rh[-r hh'(4r^2+g)+2r^2(gh'^2-k^2\sinh^2 2\alpha)+4r^2h^2+2h^2g(-2+r^2\alpha'^2)]\nonumber\\
&-2i\omega h_{2\,3} k h^2(2h+rh') \sinh 2\alpha\Big)=0\,,
\end{align}
and a first-order constraint equation
\begin{align}\label{constr2}
i\omega r^3h^3 \left(\frac{h_{t\,x}}{h^2}\right)'+2kr^3f^2gh\sinh^22\alpha\left(\frac{h_{23}}{r^2\sinh 2\alpha}\right)'=0\,.
\end{align}
One can check that the constraint equation combined with either one of the second-order equations implies
the other second-order equation.

This set of coupled linear ODEs is to be solved numerically  subject to 
the following boundary conditions. At the black hole horizon we 
demand in-going boundary conditions \cite{Son:2002sd} and choose:
\begin{align}\label{bhing}
h_{t x_1}&=(r-r_+)^{1-\frac{i\omega}{4 \pi T}} [h_{t x_1}^{(+)}+\mathcal{O}((r-r_+))]\,,\nonumber\\
h_{2 \,3}&=(r-r_+)^{-\frac{i\omega}{4 \pi T}} [h_{2\, 3}^{(+)}+\mathcal{O}((r-r_+))]\,.
\end{align}
 Using the equations of motion we find that this expansion is fixed by only one parameter $h_{t x_1}^{(+)}$ with
\begin{equation}
h_{2\, 3}^{(+)}=\frac{r_+ (4 f_+ r_+ - i \omega)}{4 (e^{2 \alpha_+} -e^{-2 \alpha_+}) f_+^2 k} h_{t x_1}^{(+)}\,.
\end{equation}
In the UV, as $r\to\infty$, we impose the following expansion
\begin{align}\label{eq:UVexpper}
h_{t x_1}&=r^2s_1+\frac{i\omega k}{2} \sinh{2\adef}s_2+\frac{v_{1}}{r^2}+ \cdots,\notag\\
 h_{2\,3}&=r^2s_2+(\frac{1}{2}i \omega k \sinh{2\adef}s_1+\frac{\omega^2}{4}s_2-k^2 \cosh^2{2\adef}s_2)+\frac{v_2}{r^2}+\dots\,,
\end{align}
where the dots include terms involving logarithms. The constraint \eqref{constr2} implies the following relation for the UV data $s_i$, $v_i$:
\begin{align}
\label{eq:constrUV}
&64i \omega v_{1}+128 k  \sinh{2 \adef}v_{2}+i\omega(-128 c_h  + 16 k^4 \sinh{2\adef}^4) s_1\nonumber\\
& -4 k \left(64 c_\alpha \cosh{2 \adef}+4k^2 \sinh{2\adef}^3(2 k^2-\omega^2+2k^2\cosh{4\adef})\right)s_2=0\,.
\end{align}

As already mentioned, it is sufficient to solve the first-order constraint equation \eqref{constr2} combined
with the second-order equation for $h_{23}$, and hence a solution to the equations of motion is specified in terms of three integration constants. On the other hand, the UV and IR expansions are specified in terms of $h_{t x_1}^{(+)}, s_{_1},s_{2}, v_{1}$, with $v_{2}$ (say) determined from this data via the
constraint. Since the ODEs we want to solve are linear, we are allowed to rescale one of the constants to unity; we choose to set $h_{t x_1}^{(+)}=1$. Thus, we are left with three nontrivial pieces of UV and IR data, which matches the number of integration constants in the problem.

\subsection{The Green's function}
The two-point function matrix, ${G}$, is defined as 
\begin{equation}
\label{eq:currents}
J_i={G}_{ij} s_j\,,
\end{equation}
where  $J_i$ are the linear-response currents that are generated by the sources $s_i$. Here the currents are the stress-tensor components
defined, as usual, as the on-shell variations\footnote{Note that throughout we write e.g. $\delta h_{x_1 t}=\delta h_{tx_1}$, so that we have
$\frac{1}{2}T^{\mu\nu}\delta h_{\mu\nu}=T^{tx_1}\delta h_{tx_1}+\dots$. This accounts
for the absence of the usual factor of 2 in \eqref{eq:defcur}.}
\begin{align}
\label{eq:defcur}
&J_1=\langle T^{t\,x_1}\rangle=\lim_{r\to\infty} \frac{1}{\sqrt{-g_\infty}} \frac{\delta S^{(2)}}{\delta h_{t\,x_1}(r)}\,,\nonumber\\
&J_2=\langle T^{\omega_2\,\omega_3}\rangle=\lim_{r\to\infty} \frac{1}{\sqrt{-g_\infty}} \frac{\delta S^{(2)}}{\delta h_{2\,3}(r)}.
\end{align}

To calculate the currents as a function of the sources, we consider a variation of the action that is quadratic in the perturbation and then put it on-shell. 
Doing so, after discarding some total derivatives in time, we find\begin{align}\label{eq:delS2}
 \delta S^{(2)}=&\int d rd^2{x}\frac{d\omega}{2\pi}\Big(\frac{r^2} { f h}(\frac{g'}{g}+2\frac{f'}{f}+2\frac{h'}{h})h_{t x_1}\delta h_{t x_1}
 -\frac{ r^2}{2 f h}( h_{t x_1}'\delta h_{t x_1}+2h_{t x_1} \delta h_{t x_1}') \nn
 &-\frac{4 g h f}{ r^3}h_{23}\delta h_{23}+\frac{ fgh}{ r^2} ( h_{23}'\delta h_{23}+2h_{23} \delta h_{23}')\Big)'
+c.t.+log\,,
\end{align}
where, for simplicity, 
we have not written out the contributions from the counterterms ($c.t.$)
and log terms ($log$) (the Minkowski analogues of \eqref{ctermp}, \eqref{ctermlog}),
which certainly play a key role,
and e.g. $h_{t x_1}\delta h_{t x_1}=h_{t x_1}(\omega)\delta h_{t x_1}(-\omega)$.
To obtain this expression we have substituted in \eqref{fourts}
and carried out the integral over time as well as over one of the $\omega$'s. We next observe that the total derivative in $r$ picks up contributions from the UV boundary and also, in principle, from the black hole horizon.
We assume that the variation
obeys the in-going boundary conditions at the black hole horizon analogous to \eqref{bhing}. 
We then observe that since both $g(r_+)$ and $h_{t x_1}(r_+)$ vanish, 
there is only a potential contribution from the horizon from the last two terms.
Inspired by \cite{Son:2002sd} we discard these terms.

To proceed we take the variations of e.g. $h_{tx_1}(\omega,r)$ and $\bar h_{tx_1}(\omega,r)$ to be independent, with $\omega\ge0$ (see \eqref{reality}).
More specifically, we are interested in variations with respect to the sources $\{\delta s_i(\omega),\delta \bar s_i(\omega)\}$.
In fact, we can deduce that these are indeed the source terms by also allowing for variations of $\delta v_{i}(\omega), \delta \bar v_{i}(\omega)$
and showing that the latter variations drop out.
After some calculation we find
\begin{align}\label{varnact1}
\delta S^{(2)}_\infty=&\int d^2 x\int_{\omega\ge 0}\frac{d\omega}{2\pi}\Big(\delta \bar s_i(\omega) J_i(\omega)+\delta s_i(\omega) \bar J_i(\omega)\Bigg)\,,
\end{align}
now integrating just over\footnote{Note that there is just a single term for $\omega=0$.} $\omega\ge 0$, and
\begin{align}
\label{eq:J1J21}
J_1=&s_1(\frac{3}{32} k^4 - 3 M  + \frac{3}{8} k^2 \omega^2 - 
 \frac{11}{24}  k^4 \cosh{4 \adef} -  \frac{3}{8} k^2 \omega^2 \cosh{4 \adef}+  \frac{35}{96} k^4 \cosh{8 \adef} )\nonumber\\
      &\qquad -s_2\,\frac{i\omega k}{8}(10k^2 -3\omega^2+14k^2 \cosh{4\adef})\sinh{2\adef}- 4 v_{1}\,,\nonumber\\
 J_2=&s_2(-M-\frac{3}{32}  k^4 + \frac{3}{8} k^2 \omega^2 - 
 \frac{3}{16} \omega^4 - \frac{37}{24} k^4 \cosh{4 \adef} + 
 \frac{9}{8} k^2 \omega^2 \cosh{4 \adef}- 
\frac{ 131}{96} k^4 \cosh{8 \adef})\nonumber\\
 &\quad+s_1\,\frac{3i\omega k}{8}(2k^2 - \omega^2+6k^2 \cosh4 \adef) \sinh2 \adef + 4v_{2}\,,
\end{align}
where $\bar J_i$ are the complex conjugate of the $J_i$.
At this stage one can use the constraint \eqref{eq:constrUV} to remove one of the $v_i$'s and continue with the calculation, however 
we choose not to impose the constraint until the very end of the calculation as this keeps the formulae more symmetric. 

To obtain the two-point function matrix, ${G}_{ij}$, we now want to 
differentiate the expression for the currents, $J_i$, with respect to the sources $s_j$ via
\begin{align}\label{geedef}
G_{ij}=\partial_{s_j}J_i.
\end{align}
Taking the derivatives of the $v_i$ with respect to the $s_j$ is actually a bit subtle since,
following the discussion at the end of section \ref{thepert}, in the gauge we are using, we cannot independently
vary the sources $s_i$. The resolution is to utilise gauge transformations and a key observation is that
there is a residual gauge freedom that acts on the boundary data. Specifically,
if we consider the coordinate transformation on the background solution given by
\begin{align}\label{gtone}
x_1\to x_1+e^{-i \omega t}\epsilon_0\,,
\end{align}
where the constant\footnote{In
Appendix \ref{altder} we will return to the fact that a constant gauge transformation
violates the in-going boundary conditions
at the black hole horizon.} $\epsilon_0$ is of the same order as the perturbation,
then this induces the residual gauge transformation acting on the metric perturbations:
\begin{align}
\label{eq:gauge0}
h_{t x_1}(r,\omega)&\to h_{t x_1}-i\epsilon_0\omega h^2 \,,\nn
h_{2\,3}(r,\omega)& \to h_{2\,3}-2\epsilon_0\, k r^2 \sinh{2\alpha}\,.
\end{align}
By expanding at the $AdS$ boundary we find that this induces the transformations on the UV data
\begin{align}
\label{gtexp}
s_1&\to s_1-i\epsilon_0 \omega,\quad\quad\quad\qquad v_{1}\to v_{1}-2i\epsilon_0\omega(c_h+\frac{k^4}{8}\sinh^4{2\adef})\,, \nn
s_2&\to s_2-2\epsilon_0 k \sinh{2\adef},\quad  v_{2}\to v_{2}-\epsilon_0 k \cosh 2\adef(4c_\alpha+k^4\cosh2\adef \sinh^32\adef)\,.
\end{align}
One can check that the constraint \eqref{eq:constrUV} is consistent with these transformations.
In Appendix \ref{altder} we will show that these gauge transformations imply that the correct derivatives
that should be used are given by
\begin{align}\label{lastone}
&\partial_{s_1} v_1=\frac{(2 k \sinh{2 \adef})v_1-i\omega (2 c_h + \tfrac {k^4} {4}\sinh^4 {2\adef})s_2}{2 k \sinh{2 \adef} s_1-i\omega s_2}\,,\nonumber\\
&\partial_{s_1} v_2=\frac{(2 k \sinh{2 \adef})v_2-k \cosh {2 \adef} (4  c_\alpha+ \, k^4 \cosh {2 \adef} \sinh^3 {2 \adef})s_2}{2 k \sinh{2 \adef} s_1-i \omega s_2}\,,\nonumber\\
&\partial_{s_2} v_1=\frac{i\omega (2 c_h + \tfrac {k^4} {4}\sinh^4 {2\adef}) s_1-i \omega v_1}{2 k \sinh{2 \adef} s_1-i\omega s_2}\,,\nonumber\\
&\partial_{s_2} v_2=\frac{k \cosh {2 \adef} (4  c_\alpha +  \, k^4 \cosh {2 \adef} \sinh^3 {2 \adef})s_1-i\omega v_2}{2 k \sinh{2 \adef} s_1-i\omega s_2}\,,
\end{align}
as well as $\partial_{s_i}s_j=\delta_{ij}$. The derivatives $\partial_{\bar s_i} \bar v_j$ are obtained by complex conjugation. One can check that these are consistent with the constraint \eqref{eq:constrUV}.

After some calculation, using \eqref{eq:J1J21}, \eqref{geedef}, \eqref{lastone} and \eqref{eq:constrUV}, we find that
\begin{align}\label{eq:Gij}
&G_{1\,1}(\omega)=-16 k^2 \sinh^2{2 \adef}\frac{\phi_2}{\phi_0}-T^{t\,t},\nonumber\\
&G_{2\,2}(\omega)=-4\omega^2\frac{\phi_2}{\phi_0}+\frac{1}{2 \sinh{2 \adef}}( T^{\omega_2\,\omega_2}-T^{\omega_3\,\omega_3}),\nonumber\\
&G_{1\,2}(\omega)=-G_{2\,1}(\omega)=8 i k \omega \sinh{2 \adef} \frac{\phi_2}{\phi_0}\,,
\end{align}
where $T$ refers to the background stress-tensor given in \eqref{explicstress}
and $\phi_0$ and $\phi_2$ are the following gauge-invariant combinations 
\begin{align}
\phi_0=&s_1 -i\frac{s_2\omega }{2 k \sinh 2 \adef}\,,\nonumber\\
\phi_2=&\frac{v_1}{4 k^2 \sinh{2 \adef}^2}-\frac{s_1}{64 k^2}(8 k^4-3 k^2\omega^2+16k^4 \cosh{4\adef}+\frac{32c_h}{\sinh^2{2 \adef}})\nonumber\\
&+i\frac{s_2\omega}{128\,k \sinh{2 \adef} }(10 k^2-3 \omega^2+14k^2\cosh{4\adef})\,.
\end{align}
Note that $G_{12}(\omega)=-G_{21}(\omega)$
is expected since the deformation does not break time-reversal invariance
and $T^{tx_1}$ and $T^{\omega_2\omega_3}$ are odd and even operators under time-reversal, respectively.
In Appendix \ref{altder}, as an aside, we will show that by taking two derivatives of the on-shell action one does
not recover the Green's function but rather the Hermitian combination $G+G^\dagger$.

Let us now consider the Green's functions in the limit that $\omega\to 0$, which gives 
the static susceptibilities. First we observe from
\eqref{sceqs} and \eqref{constr2} that when $\omega=0$ an exact solution is given by $\delta g_{tx_1}=0$ and $\delta g_{23}=s_2 r^2\sinh 2\alpha/\sinh2\alpha_0$.
In fact this zero-mode solution is obtained from the coordinate transformation $x_1\to x_1-s_2/(2k\sinh 2\alpha_0)$ on the background solution.
Similarly, there is also another solution given by $\delta g_{tx_1}=s_1 gf^2$ with $\delta g_{23}=0$, which can be obtained
from the background solution via  $t\to t-s_1 x_1$. For both of these explicit solutions we can obtain the corresponding values of
the expectation values $v_2$ and $v_1$ as explicit functions of $s_2$ and $s_1$ for each case, respectively.
Calculating as above we deduce that\footnote{It is interesting to compare our results to that of AdS-Schwarzschild. Carrying out the above derivation we
find that $G_{11}(\omega)=-3M=-T^{tt}$, $G_{12}(\omega)=0$ and $G_{22}(\omega)=-M-3 \omega^4/16+4v_2/s_2$. Note that
for $G_{11}$ there is also a hidden delta function.}
\begin{align}\label{sts}
\lim_{\omega\to 0}G_{11}(\omega)&= T^{x_1x_1}\,,\nn
\lim_{\omega\to 0}G_{22}(\omega)&= \frac{1}{2 \sinh{2 \adef}}( T^{\omega_2\,\omega_2}-T^{\omega_3\,\omega_3})\,,\nn
\lim_{\omega\to 0}G_{12}(\omega)&= 0\,,
\end{align}
where the stress-tensor components of the background geometry are given in \eqref{stressotherbasis}.

We now make some preliminary comments concerning the DC conductivity matrix, defined as 
\begin{align}
C_{ij}\equiv\lim_{\omega\to 0}\frac{Im G_{ij}(\omega)}{\omega}\,.
\end{align}
We will see from our numerical results in the next section that the component $C_{11}$, which fixes the DC thermal conductivity via $C_{11}=T\kappa$, 
is nonvanishing, and in section \ref{dccalc} we will obtain an analytic result in terms of black-hole horizon data. Given this, and recalling \eqref{eq:Gij}, we see that
we must have $Im(\phi_2/\phi_0)\sim\omega$ as $\omega\to 0$ and hence
we have 
\begin{align}\label{c22result}
C_{22}=0\,.
\end{align}
 On the other hand to obtain $C_{12}$ and $C_{21}$ we require the behaviour of $Re(\phi_2/\phi_0)$ as $\omega\to 0$.
This can be obtained by comparing the results \eqref{sts} with \eqref{eq:Gij} and we deduce that the off-diagonal components of the DC conductivity matrix
are given by
\begin{align}\label{c12result}
C_{12}=-C_{21}=-\frac{1}{2k\sinh 2\adef}(T^{tt}+T^{x_1x_1})\,.
\end{align}

\subsection{Numerical results}
As discussed above, it is sufficient to solve the first-order constraint equation \eqref{constr2} combined
with the second-order equation for $h_{23}$ given in \eqref{sceqs}, 
and hence a solution is specified in terms of three integration constants. 
In practice we exploit the fact that the equations are linear to set $h_{t x_1}^{(+)}=1$ and then the
three integration constants are the $s_i$ and $v_i$ subject to the constraint \eqref{eq:constrUV}.
We solved the system using a standard shooting method.

\begin{figure}[h]
\centering
{\includegraphics[width=7cm]{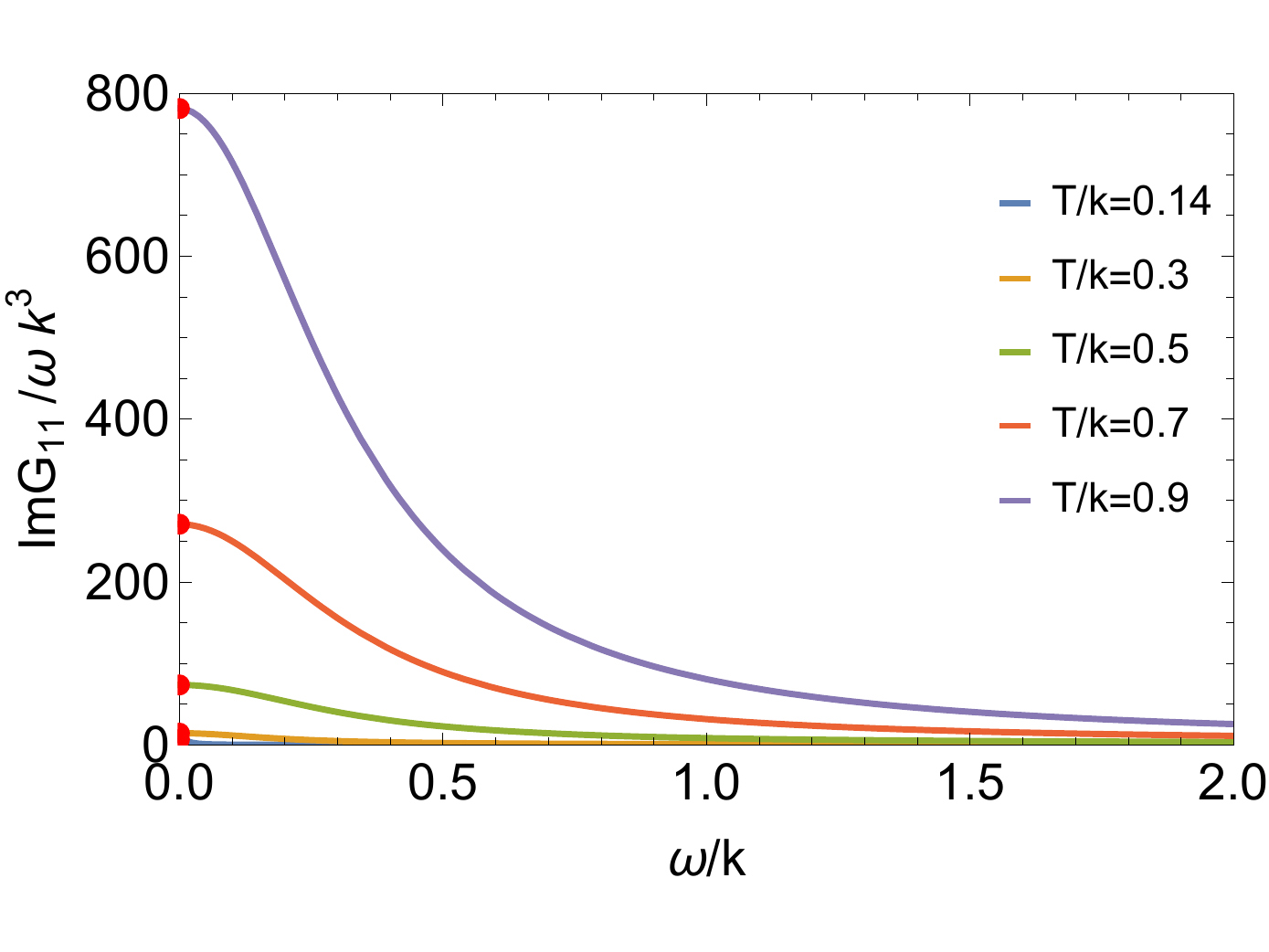}}\qquad{\includegraphics[width=7cm]{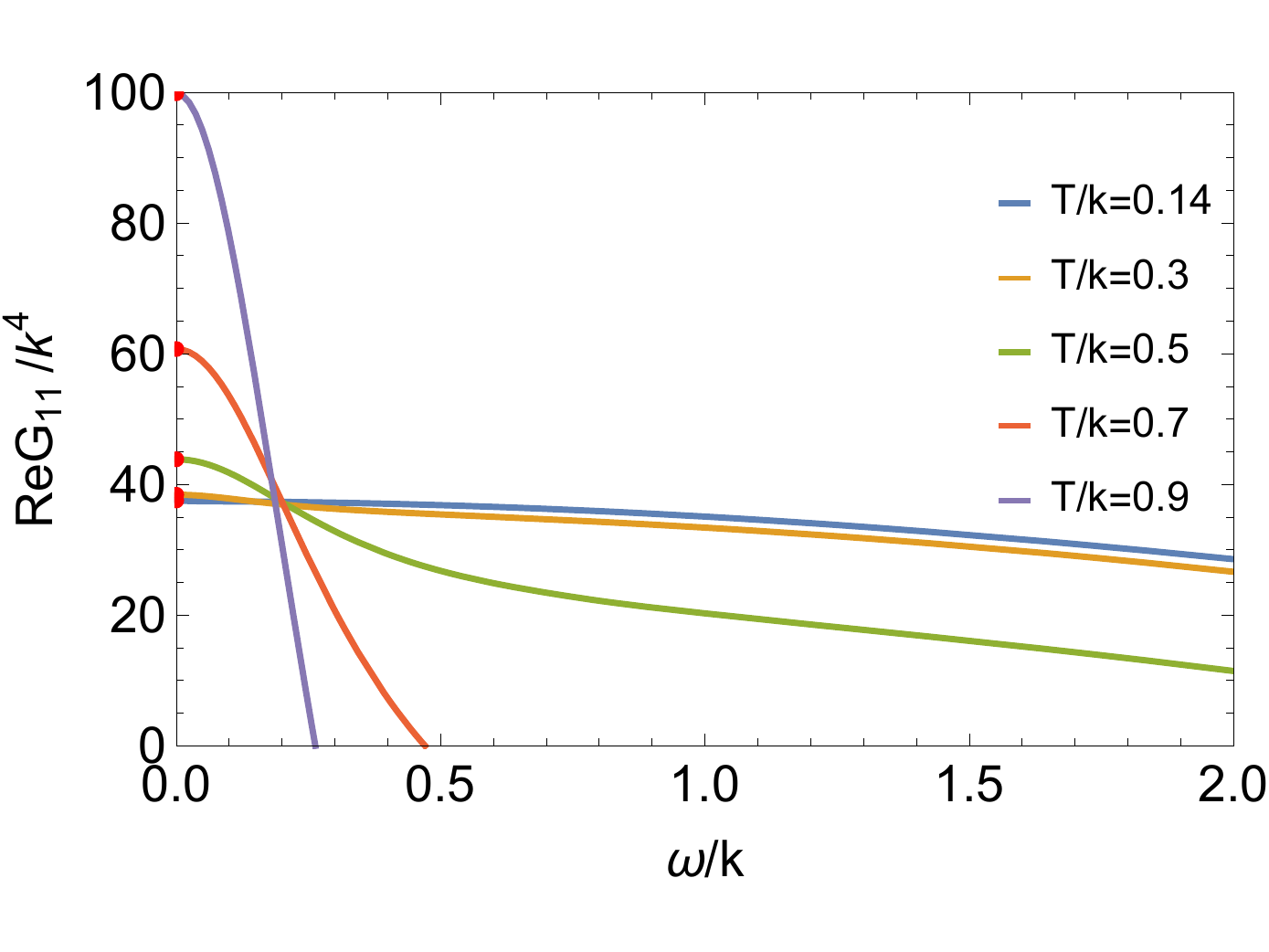}}
\caption{Plot of the real and imaginary parts of the two-point function $G_{11}\equiv\langle T^{t x_1} T^{t x_1}\rangle$, suitably scaled with $\omega$ and $k$, 
against $\omega/k$, suitably rescaled with $k$, for various values of the temperature for helical black holes with $\alpha_0=1/2$. The left panel shows the real part of the thermal conductivity,  
$Re(T\kappa(\omega))=Im(G_{11})/\omega$ with the red dots indicating the DC conductivity predicted from the results of section \ref{dccalc}. 
The right panel shows the real part of $G_{11}$ and the red dots indicate the static susceptibilities derived in \eqref{sts}.}\label{fig:3pf}
\end{figure}

In the left panel of figure \ref{fig:3pf} we have plotted the real part of the thermal conductivity $\kappa$, obtained via 
\begin{align}
T\kappa(\omega)\equiv \frac{G_{11}}{i\omega}\,,
\end{align}
for the helically deformed black holes with $\alpha_0=1/2$. Note that in this and subsequent plots, we have divided by suitable powers of $k$ to plot dimensionless quantities.
As the temperature is lowered we see the appearance of Drude-type peaks associated with the fact that
we have broken translation invariance in the $x_1$ direction. The red dots in this panel are the DC thermal conductivities
that are obtained from an analytic result in terms of black hole horizon data, which we derive in section \ref{dccalc}.
The right panel of figure \ref{fig:3pf} plots the real part of $G_{11}$.

\begin{figure}[h]
\centering
{\includegraphics[width=7cm]{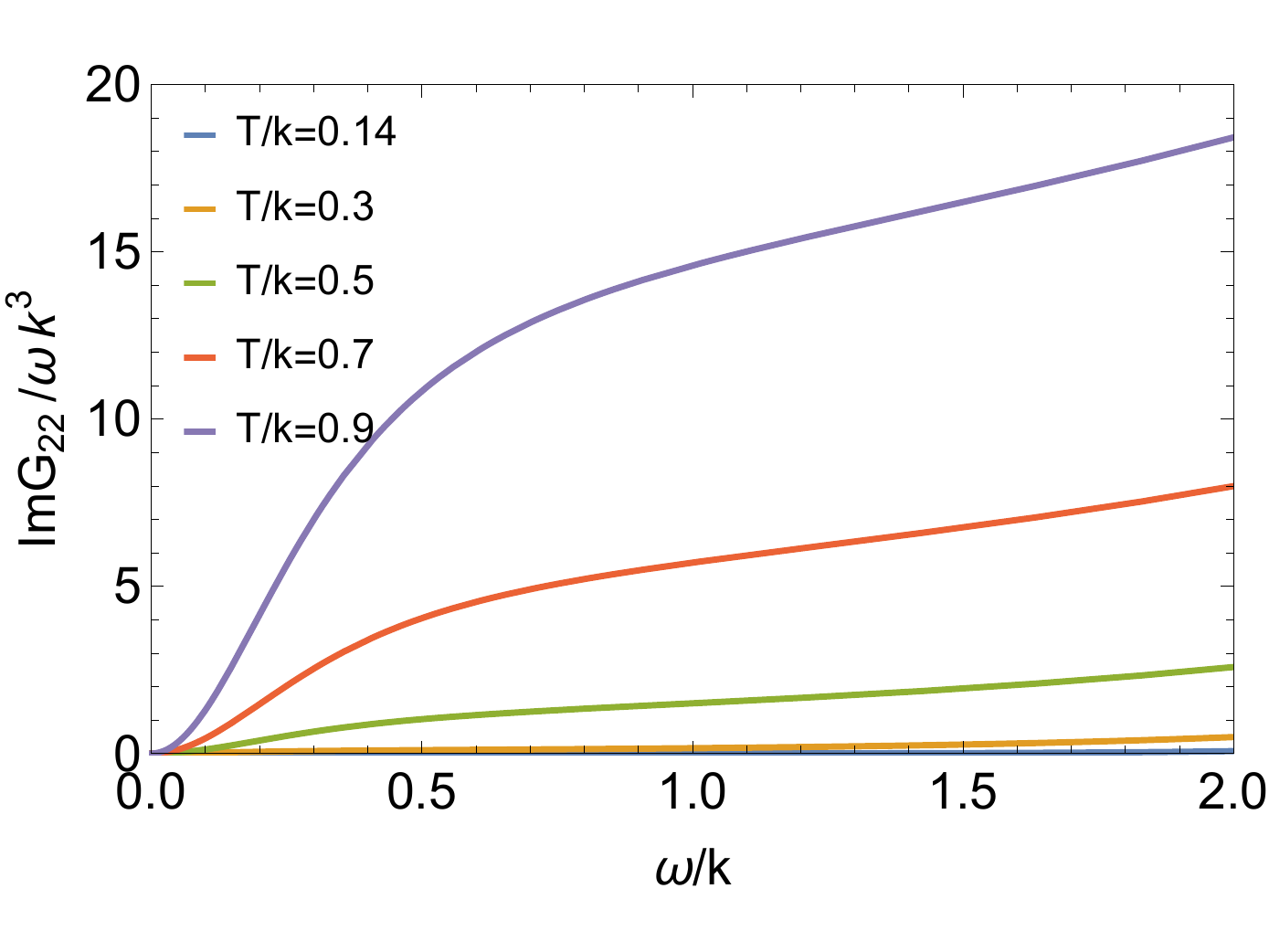}}\qquad{\includegraphics[width=7cm]{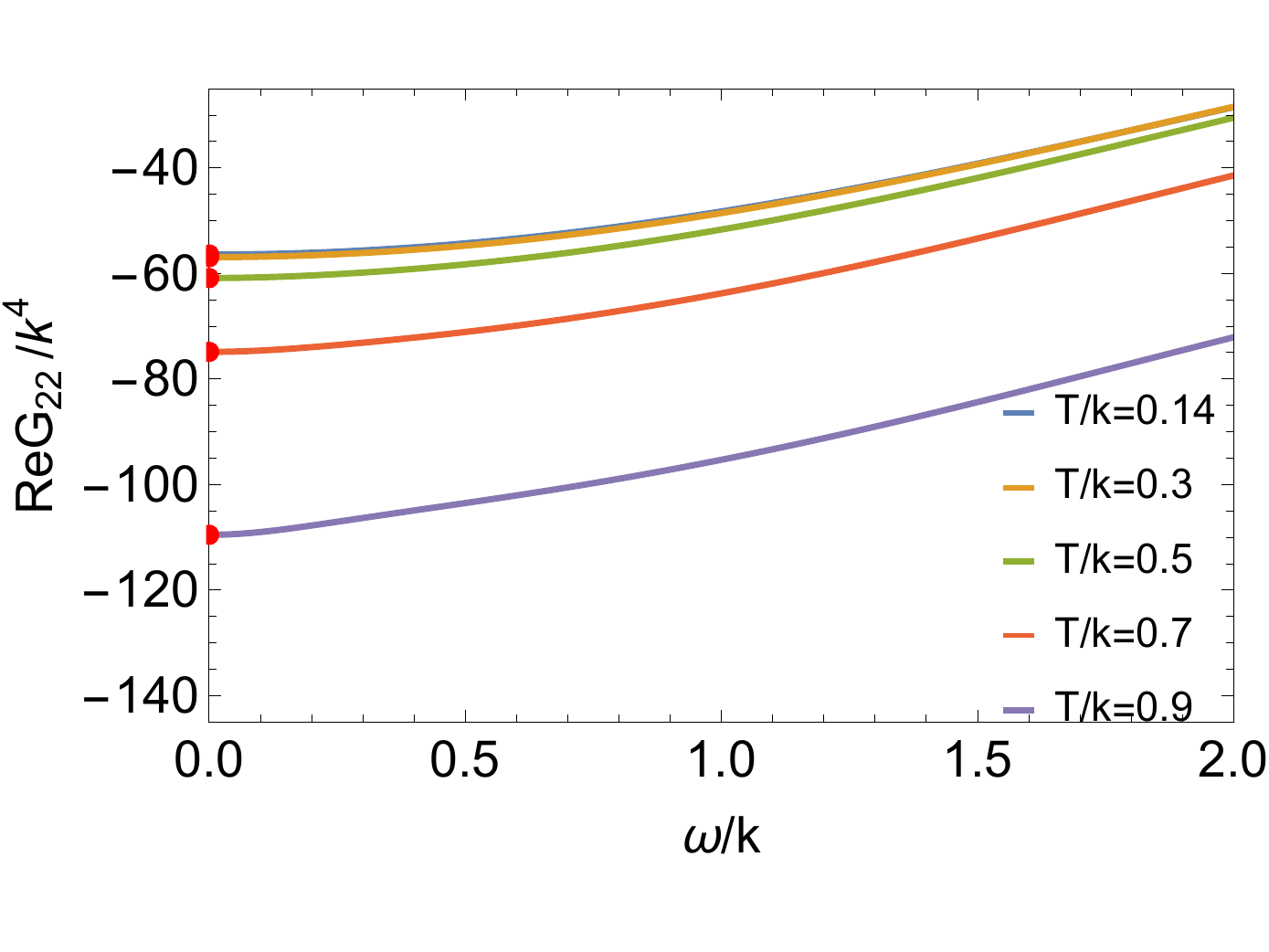}}
{\includegraphics[width=7cm]{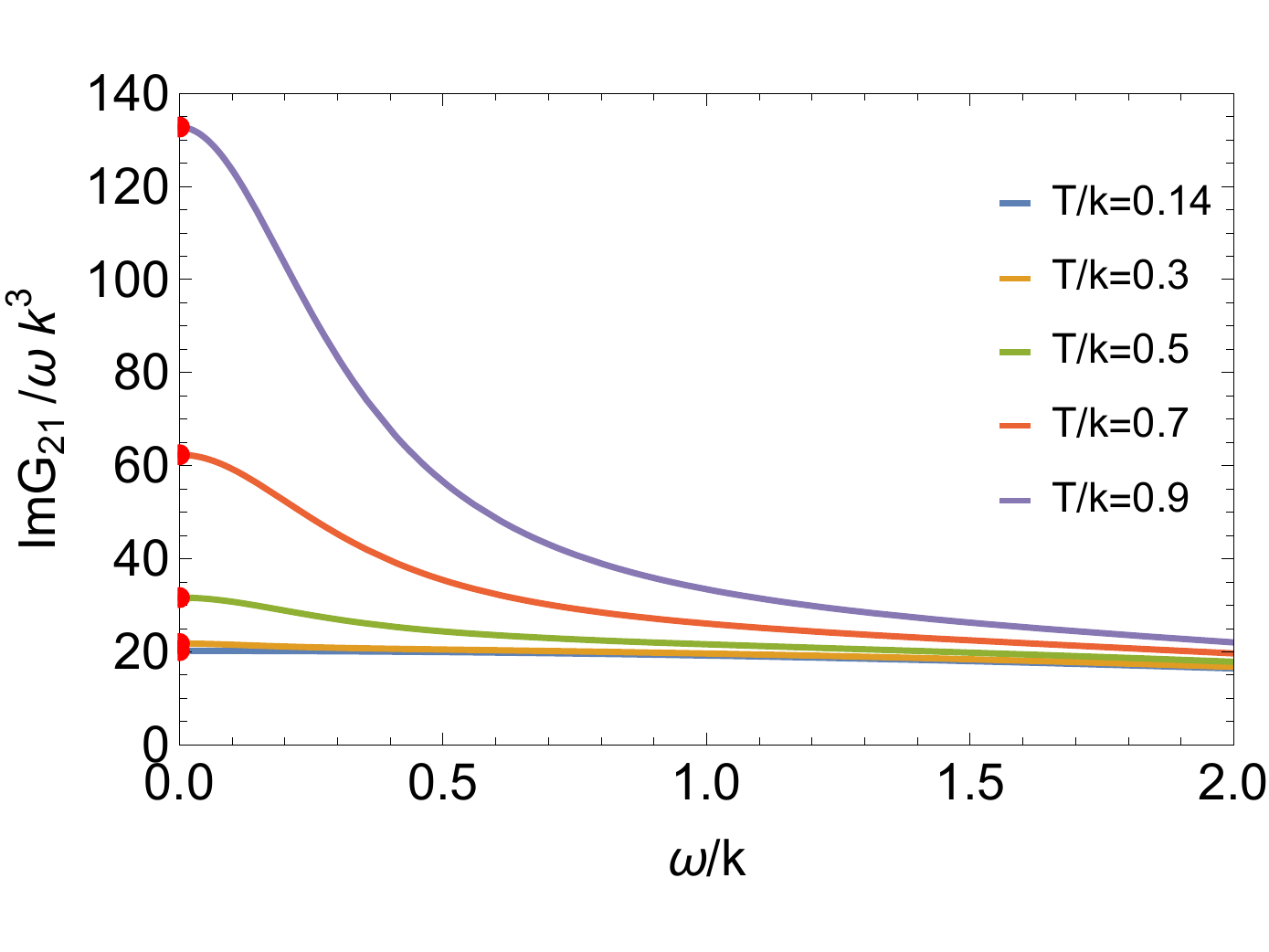}}\qquad{\includegraphics[width=7cm]{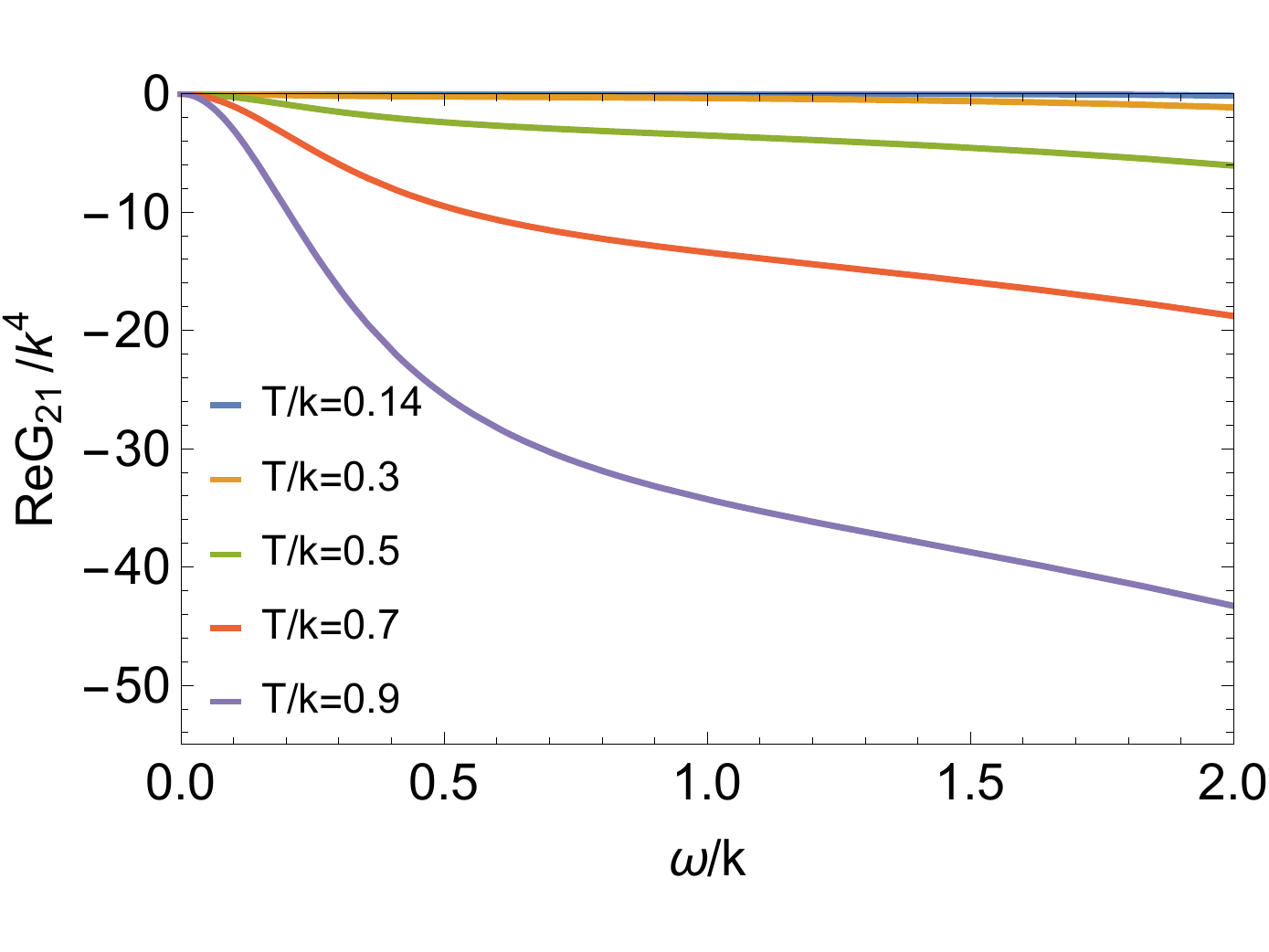}}
\caption{Plot of the real and imaginary parts of the two-point functions $G_{22}\equiv \langle T^{\omega_2\omega_3}T^{\omega_2\omega_3} \rangle$ and $G_{21}\equiv \langle T^{t x_1}T^{\omega_2\omega_3} \rangle$, suitably scaled, against $\omega/k$ for various values of the temperature for helical black holes
with $\alpha_0=1/2$. In the top left plot we see that the DC conductivity $C_{22}=0$ as in \eqref{c22result}. In the top right plot, the red dots indicate the static susceptibility derived in \eqref{sts}. In the bottom left plot, the red dots indicate the DC conductivity $C_{21}$ derived in \eqref{c12result}. In the bottom right
plot we see that the static susceptibility vanishes in agreement with \eqref{sts}; we also see that this plot is a simple rescaling of that in
the upper left plot, as expected from \eqref{eq:Gij}.}\label{fig:4pf}
\end{figure}

In figure \ref{fig:4pf} we present the corresponding plots for $G_{21}$ and $G_{22}$ for the same
background black holes with $\alpha_0=1/2$. We observe the Drude peaks in $Im(G_{21})/\omega$ which, from \eqref{eq:Gij}, have the same origin
as the Drude peaks in $Im(G_{11})/\omega$.

\section{DC thermal conductivity from the black hole horizon}\label{dccalc}
In this section we will derive an expression for the thermal DC conductivity $\kappa\equiv \lim_{\omega\to 0}\kappa(\omega)$
in terms of black hole horizon data, following the approach of \cite{Donos:2014uba,Donos:2014cya}. The final result is given in \eqref{kapres}.
Recall that $\kappa=C_{11}/T$. We will also recover our previous results for the other DC conductivity matrix elements $C_{22}$, and $C_{12}$, $C_{21}$ given in \eqref{c22result} and \eqref{c12result}, respectively, as well as the static susceptibilities $G_{ij}(\omega=0)$ given in \eqref{sts}.

As explained in \cite{Donos:2014uba,Donos:2014cya} the strategy is to switch on sources for the operators
$T^{tx_1}$ and  $T^{\omega_2\omega_3}$ that are linear in time, $s_i=b_i t$, where $b_i$ are constant parameters,
and then read off the linear response. As shown in Appendix C of \cite{Donos:2014cya}, the expectation values of the
operators will then be given by
\begin{align}\label{teeeqnst}
T^{tx_1}(t)=\left[t G_{1j}(\omega=0)-C_{1j}\right]b_j\,,\nn
T^{\omega_2\omega_3}(t)=\left[t G_{2j}(\omega=0)-C_{2j}\right]b_j\,,
\end{align}
and hence, given the expectation values, we can deduce the DC conductivity matrix, $C_{ij}$, as well as 
the static susceptibility matrix, $G_{ij}(\omega=0)$.

\subsection{Linear in time source for $T^{tx_1}$}
Following the construction of \cite{Donos:2014uba,Donos:2014cya}, we consider perturbations around the black holes of section \ref{sec:BH} of the form
\begin{align}
&g_{t x_1}(t,r)=-\zeta\, t \delta F(r)+h_{t x_1}(r)\,,\quad
g_{r x_1}(r)=h_{r x_1}(r) \,,\quad
g_{\omega_2\omega_3}(r)= h_{23}(r)\,,
\end{align} 
where $\zeta$ is a constant.
It is straightforward to show that the linearised Einstein equations reduce to
one equation that can be algebraically solved for $h_{r x_1}$ in terms of $h_{23}$, $h_{23}'$:
  \begin{align}
\label{eq:hrx}
h_{r x_1}&=\frac{ \zeta h}{2 k^2 r g (2 h + r h')\sinh^2 2 \alpha}\Big( -6  r^2 h^2 +   g h^2 +  k^2 r^2 \sinh^2{2 \alpha} + 4  r g h h'  +  r^2 g h'^2-   r^2 g h^2 \alpha'^2
   \Big)\nn
&-\frac{h^2}{2k}\left(\frac{h_{23}}{r^2\sinh2\alpha}\right)'\,,
  \end{align}
a second-order ODE for $h_{t x_1}$:
  \begin{align}
  \label{eq:eomDC}
&h_{tx_1}''=\frac{1}{r g h^2 (2 h + r h')} \Big(
h_{tx_1}' h[ 2 r^2 g h'^2 + 
  2 r^2 g  h^2  \alpha'^2+ 4 r^2 h^2 -  4 g h^2 
\nn
&\qquad\qquad\qquad\qquad\qquad \qquad\qquad-4r^3hh'-rghh' -   2 k^2 r^2  \sinh^2{2 \alpha} ]\nn
&+h_{tx_1}4[4 r  h^3    - 4 r^2 h^2 h' + 2 r g  h h'^2 -  r^2 g h^2 h' \alpha'^2
+   g  h^2 h'+   k^2 r^2  \sinh^2{2 \alpha} h'  ]\Big)\,,
\end{align}
as well as a second-order ODE for $\delta F$ which can be integrated to give 
\begin{equation}\label{effsol}
\delta F=f^2g\,.
\end{equation}
Now, following the same discussion as in \cite{Donos:2014cya}, with this $\delta F$
we deduce that
\begin{align}\label{beedef}
b_1\equiv-\zeta\,,
\end{align}
is parametrising a time-dependent source for the heat current $T^{tx_1}$.

We next obtain a first integral for the equation of motion for $h_{t\, x_1}$. To do so we
consider the two-tensor
\begin{equation}
G^{\mu\nu}=\nabla^\mu k^\nu\,,
\end{equation}
where $k=\partial_t$. Using the equations of motion we can show that $\partial_r(\sqrt{-g}G^{r x_1})=0$ and thus we can conclude that 
\begin{align}\label{Qexp}
Q=2\sqrt{-g}G^{r x_1}&=\frac{r^2 f^3 g^2}{h}\partial_r\left(\frac{h_{t\, x_1}}{g f^2}\right)\,,
\end{align}
 is a constant and hence can be evaluated at any value of $r$. Evaluating it at $r\to \infty$ we
 will now show that $Q$ is the time-independent part of the heat current, $Q=T^{tx_1}_0$. 
 By evaluating $Q$ at the horizon, after ensuring the perturbation is regular at the horizon, will lead to the final
 expression for the thermal DC conductivity $\kappa$.

Using \eqref{stressy} to calculate the stress tensor component $\tilde{T}^{t\, x_1}$ for the perturbed metric, at a general value of $r$ and to first order in the perturbation, we can show that
\begin{align}
 Q=\frac{r^2 f^3 g^2}{h}\partial_r\left(\frac{h_{t\, x_1}}{g f^2}\right)
 =r^2 f h\sqrt{g}\,(f^2 g\,\tilde{T}^{t\, x_1}-( h_{t x_1}-\zeta t g f^2) \tilde{T}^{x_1\,x_1})\,,
 \end{align}
where $\tilde{T}^{x_1\,x_1}$ is a component of the stress tensor of the background given in \eqref{stressotherbasis}.
Since $Q$ is time-independent, the time dependent piece of $\tilde T^{t\, x_1}$ must cancel with the time dependent piece coming from the second term. In other words, 
\begin{equation}\label{eq:tdef}
\tilde{T}^{\,t\, x_1}\equiv    \tilde{ T}^{\,t\, x_1}_{0}  -    \zeta t \tilde{T}^{\,x_1\,x_1}\,,
\end{equation}
where $ \tilde{ T}^{\,t\, x_1}_{0}$ is time-independent, and hence
 \begin{equation}\label{eq:J}
Q=r^2 f h\sqrt{g}\,(f^2 g\,\tilde{T}^{t\, x_1}_0-h_{t x_1} \tilde{T}^{x_1\,x_1})\,.
 \end{equation}
We will demand that $h_{t x_1}\sim r^{-2}$ close to the boundary and hence the first term in the brackets dominates the second term and
we conclude that at $r\to \infty$ we have $Q=r^6\tilde{T}^{t\, x_1}_0={T}^{t\, x_1}_0$ as claimed. Thus,
we have
\begin{align}
{T}^{\,t\, x_1}=    Q -    \zeta t {T}^{\,x_1\,x_1}\,.
\end{align}
At this point, using \eqref{teeeqnst} and recalling \eqref{beedef}, we see that the explicit time dependence implies that 
$G_{11}(\omega=0)= T^{x_1x_1}$, in agreement with
the static susceptibility derived earlier in \eqref{sts}. 

To evaluate $Q$ at the black hole horizon we need to know the behaviour of $h_{t x_1}$ as $r\to r_+$. 
Allowing $h_{2\,3}$ to be constant at the horizon, using equation \eqref{eq:hrx} we see that $h_{r x_1}$ diverges at the horizon as
\begin{align}\label{hexpr}
h_{r x_1}=-\frac{\zeta h_+^2}{4k^2\sinh^2 2\alpha_+}\frac{1}{(r-r_+)}+\dots \,.
\end{align}
Notice that $h_{2\,3}$ is not constrained in any other way; we choose it so that 
$h_{2\,3}$ and also $h_{rx_1}$ fall-off fast enough as $r\to \infty$ so that they do not contribute
to any source as $r\to\infty$; we will return to this point below.
Now, given \eqref{hexpr} and \eqref{effsol}, we ensure that the perturbation is regular at the horizon by using
in-going Eddington Finklestein coordinates $(v,r)$, where
$v=t+\log(r-r_+)^{\frac{1}{g_+f_+}}$, and we deduce that 
the behaviour of $h_{t x_1}$ should be
\begin{equation}
h_{t x_1}\sim  \left. f_+ g_+(r-r_+) h_{r x_1} \right|_{r=r_+} -  {\zeta f_+ (r-r_+) }  \ln(r-r_+)+\cdots\,.
\end{equation}
Importantly, one can check that this expansion can also be obtained directly from the near horizon expansion of the differential equation for
$h_{tx_1}$ in \eqref{eq:eomDC}. In fact this expansion imposes only a single condition at this boundary and, as we mentioned above,
we impose that as $r\to \infty$ we have the behaviour $h_{tx_1}\sim r^{-2}$. Together these two conditions give a unique solution to the differential equation
in \eqref{eq:eomDC}.
Having obtained a regular perturbation we can now use \eqref{Qexp} to obtain an expression for $Q$ evaluated at the horizon. Using \eqref{teeeqnst} we have $C_{11}=-T^{tx_1}/b_1=-Q/b_1$ and since $C_{11}=T\kappa$,
we
deduce the following expression for the thermal conductivity $\kappa$ in terms of horizon data:
\begin{align}\label{kapres}
\kappa
= \frac{\pi s T}{k^2 \sinh^2 2\alpha_+ }\,.
\end{align}

For the black hole backgrounds that we constructed explicitly in section \ref{sec:BH}, we have checked that this result agrees precisely
with the $\omega\to 0$ limit of the AC conductivity. This is displayed for a particular helical deformation, for various
temperatures, in figure \ref{fig:3pf}.
We can also use the analytic result \eqref{kapres} to obtain the low-temperature behaviour of the thermal conductivity for the helically deformed black holes. Indeed, following
the analysis leading to \eqref{scess}, we find that for $T<<0$ we have the leading-order behaviour:
\begin{align}
\kappa\sim 
\left(\frac{\pi^8\bar h_+}{\bar\alpha_+^2\bar f_+^6k^5}\right)T^7e^{4 \bar f_+ k/\bar h_+ \pi T}\,.
\end{align}

We now return to a point mentioned above. Consistent with \eqref{eq:hrx} we choose the asymptotic expansion of $h_{23}$ as $r\to\infty$
to be given by
\begin{align}\label{h23finex}
h_{23}=\zeta\left(\frac{k\sinh 2\adef }{2} + \frac{32 c_h+8M-k^4+k^4 \cosh8 \adef}{16 k \sinh 2 \adef r^2} - \frac{k^3 \sinh 6\adef \log r}{3r^2}+\dots \right)\,,
\end{align}
which ensures that the $1/r$, $1/r^3$ and $\log r/r^3$ terms in the asymptotic expansion of $h_{rx_1}$ all vanish. 
It is clear from \eqref{h23finex} that the perturbation that we are considering does not have a non-normalisable source term, as claimed above.
However, there is a corresponding expectation value for $T^{\omega_2\omega_3}$. Indeed we find, using \eqref{stressy}, 
that
\begin{align}
T^{\omega_2\omega_3}=\zeta\frac{8M +32 c_h+k^4\sinh^2 2\adef(7+13 \cosh4\adef)}{4 k \sinh 2 \adef}\,.
\end{align}
Comparing with \eqref{teeeqnst} we conclude that
\begin{align}
C_{21}=\frac{T^{\omega_2\omega_3}}{\zeta}=\frac{1}{2k\sinh 2\adef}(T^{tt}+T^{x_1x_1})\,,\end{align}
as well as $G_{21}(\omega=0)=0$ in agreement with \eqref{c12result}
and \eqref{sts}, respectively.

\subsection{Linear in time source for $T^{\omega_2\omega_3}$}
We now consider perturbations around the black holes of section \ref{sec:BH} of the form
\begin{align}
&g_{t x_1}(r)=h_{t x_1}(r)\,,\qquad
g_{r x_1}(r)=h_{r x_1}(r) \,,\qquad
g_{\omega_2\omega_3}(t,r)= \cdef t \delta F(r) + h_{23}(r)\,,
\end{align}
where $\cdef$ is a constant.
After substituting in the equations of motion we find that it is consistent to take
\begin{align}
\delta F =r^2\sinh2\alpha\,.
\end{align}
Note, for later use, that since $\alpha\to\adef$ at $r\to\infty$, the source is parametrised by
\begin{align}
b_2\equiv \cdef\sinh2\adef\,.
\end{align} 
We also find that we can solve for $h_{rx_1}$ algebraically:
\begin{align}\label{grxeqn2}
h_{rx_1}=-\frac{h^2}{2k}\partial_r\frac{h_{23}}{r^2\sinh2\alpha}\,,
\end{align}
and we can also obtain a second order differential equation for $h_{tx_1}$ which, remarkably,
we can cast in the form
\begin{align}
\partial_r\tilde Q=0\,,
\end{align}
where 
\begin{align}
\tilde Q=Q-\frac{\cdef }{k}rfg(h-rh')\,,
\end{align}
and $Q$ is given in \eqref{Qexp}.

We next examine the regularity of the metric at the horizon. Considering
$g_{\omega_2\omega_3}$ and using Eddington-Finklestein coordinates we must have
$h_{23}\sim \frac{\cdef r_+^2\sinh2\alpha_+}{g_+f_+}\log(r-r_+)$. Then using \eqref{grxeqn2}
we can deduce that $h_{rx_1}\sim -\frac{\cdef h_+^2}{2k g_+ f_+}\frac{1}{(r-r_+)}$. Again using Eddington-Finklestein coordinates, this behaviour
of $h_{rx_1}$ at the horizon implies that $h_{tx_1}\sim h_{tx_1}^+$ with 
\begin{align}\label{hplusexp}
h_{tx_1}^+=-\frac{\cdef h_+^2}{2k}\,.
\end{align}

We now return to the constant $\tilde Q$. Evaluating it at the horizon we have
\begin{align}
\tilde Q(r_+)&=Q(r_+)\,,\nn
&=-\frac{r_+^2f_+g_+}{h_+}h_{tx_1}^+\,,\nn
&=\cdef \frac{Ts}{2k}\,,\nn
&=\frac{\cdef }{2k}\left(T^{tt}+T^{x_1x_1}\right)-\frac{\cdef }{2k}\left([8c_h+2k^4\sinh^2 2\adef(1+4\cosh\adef)]\right)\,.
\end{align}
To get the first and second lines we used $g(r_+)=0$, \eqref{Qexp} and \eqref{bhas}.  
To get the third line we used \eqref{hplusexp}, while the last line is obtained using
the Smarr-type formula \eqref{smarrp} as well as
\eqref{explicstress}.
On the other hand, evaluating at $r\to \infty$ we first find, using \eqref{stressy}, that
\begin{align}\label{qtill}
\tilde Q= r^2 f h\sqrt{g}\,\left(f^2 g\,\tilde{T}^{t\, x_1}- h_{t x_1} \tilde{T}^{x_1\,x_1}+
\cdef \left(\frac{k\sinh^22\alpha}{h^2}-\frac{g^{1/2}(h-rh')}{krh}\right)\right)\,,
\end{align}
As $r\to\infty$ we find that the first and last terms give a contribution leading to
\begin{align}
\tilde Q={T}^{t\, x_1}
-\frac{\cdef }{2k}\left([8c_h+2k^4\sinh^2 2\adef(1+4\cosh\adef)]\right)\,,
\end{align}
and we thus have
\begin{align}
{T}^{t\, x_1}=\frac{\cdef }{2k}\left(T^{tt}+T^{x_1x_1}\right)\,.
\end{align}
Using \eqref{teeeqnst} we now deduce that
\begin{align}
C_{12}=-\frac{{T}^{t\, x_1}}{\cdef \sinh 2\adef}=-\frac{1}{2k\sinh 2\adef}(T^{tt}+T^{x_1x_1})\,,
\end{align}
in agreement with \eqref{c12result}, as well as $G_{12}(\omega=0)=0$, in agreement with \eqref{sts}.

Finally, returning to \eqref{grxeqn2} and demanding that the $1/r$, $1/r^3$ and $\log r/r^3$ terms in the asymptotic expansion of $h_{rx_1}$ all vanish we deduce that the constant, $1/r^2$ and $\log r/r^2$ terms 
of $h_{23}$ all vanish. Using \eqref{stressy} we then find
\begin{align}
T^{\omega_2\omega_3}=\frac{\cdef}{2}  t ( T^{\omega_2\,\omega_2}-T^{\omega_3\omega_3})\,.
\end{align}
Using \eqref{teeeqnst} we thus recover the result \eqref{c22result} that $C_{22}=0$ and moreover  
$G_{22}(\omega=0)=\frac{1}{2 \sinh{2 \adef}}( T^{\omega_2\,\omega_2}-T^{\omega_3\,\omega_3})$.

\section{Final Comments}
Using holographic techniques we have analysed in some detail a universal helical deformation that all
$d=4$ CFTs possess. The deformation is specified by a wave number $k$, the strength of the deformation $\alpha_0$
and a dynamical scale that is introduced due to the conformal anomaly.
We constructed black hole solutions that describe the deformed CFTs at finite temperature
for a range of $k,\alpha_0$. By analysing the low-temperature behaviour of the black hole solutions we showed that
the deformed CFTs approach, in the far IR, the undeformed UV CFTs, up to a renormalisation of length scales.
This is similar to what was seen in \cite{Chesler:2013qla} for the deformation of $d=3$ CFTs by a periodic chemical potential which averages to zero 
over a period.

We calculated the AC thermal conductivity along the axis of the helix and showed that it exhibited Drude peaks. 
This involved a careful calculation of the two-point functions for the $T^{tx_1}$ and $T^{\omega_2\omega_3}$ components
which mix in the deformed background. We also obtained an analytic result for the DC conductivities in terms of black horizon
data, by switching on sources that are linear in time, following \cite{Donos:2014uba,Donos:2014cya}, finding a satisfying agreement with the AC results, including the static susceptibilities.

  \section*{Acknowledgements}
The work is supported by the Science and Technology Facilities Council (STFC) grant ST/J0003533/1,
the Engineering and Physical Sciences Research Council (EPSRC) programme grant EP/K034456/1,
the Generalitat de Catalunya grant 2014-SGR-1474, the Ministerio de Economia y Competitividad (MEC) grants
FPA2010-20807-C02-01 and FPA2010- 20807-C02-02,  the
Centro Nacional de F\'isica de Part\'iculas, Astropart\'iculas y Nuclear (CPAN) grant
CSD2007-00042 Consolider- Ingenio 2010
 and also by the European Research Council (ERC) under the European Union's Seventh Framework Programme (FP7/2007-2013), ERC Grant agreements STG 279943, STG 306605 and 
 ADG 339140.

\appendix
\section{Boundary energy-momentum tensor}\label{ap:StressTensor}
Here we record the explicit expressions for the components of the
energy-momentum tensor $\tilde{T}^{\mu\nu}$ defined in \eqref{stressy}. Writing
$T^{\mu\nu}=r^6\tilde{T}^{\mu\nu}$ and after setting $f_0=1$, we find
\begin{align}\label{explicstress}
T^{tt}&=3 M +8 c_h+\frac{k^4}{24}\sinh^2 2\alpha_0(35+61\cosh 4\alpha_0)\,,\nonumber\\
T^{x_1 x_1}&=M+8 c_h+\frac{k^4}{24}\sinh^2 2\alpha_0(49+95\cosh 4\alpha_0)\,,\nonumber\\
T^{x_2 x_2}&=\cosh 2\adef\left(M+8 c_\alpha \cos 2 kx_1\right)\nn
&+\sinh 2\adef\Big(        -8c_\alpha-\frac{1}{96}\cos 2kx_1\left(96 M+k^4(33+148\cosh 4\adef+107 \cosh 8\adef)\right)    \nn
&\qquad\qquad\qquad+\frac{1}{48}\sinh 4\adef k^4(-35+107\cosh 4\alpha_0)\Big)\,,
\nonumber\\
T^{x_3 x_3}&=\cosh 2\adef\left(M-8 c_\alpha \cos 2 kx_1\right)\nn
&+\sinh 2\adef\Big(        -8c_\alpha+\frac{1}{96}\cos 2kx_1\left(96 M+k^4(33+148\cosh 4\adef+107 \cosh 8\adef)\right)    \nn
&\qquad\qquad\qquad+\frac{1}{48}\sinh 4\adef k^4(-35+107\cosh 4\alpha_0)\Big)\,,\nn
T^{x_2 x_3}&=\sin{2 k x_1} \Big(-8 c_\alpha \cosh 2\adef\nonumber\\&
+\frac{\sinh2 \adef}{96}\left(96 M+k^4(33+148\cosh 4\adef+107 \cosh 8\adef)\right)\Big) \,.
\end{align} 
If we set $\adef=0$, the above agrees with the results of \cite{Donos:2013woa} in the absence of matter fields. 

We also note that we can use the results of \cite{Donos:2013cka}
to recover the two expressions for the free energy that we obtained directly in the text.
Specifically, equation (2.15) and (2.14) of \cite{Donos:2013cka} imply that
\begin{align}
w&=-Ts-\frac{k}{2\pi}\int_0^{2\pi/k}dx_1\sqrt{-\gamma}\left(\tilde T^{tt}\gamma_{ tt}\right)\,,  \nn
&=-\frac{k}{2\pi}\int_0^{2\pi/k}dx_1\sqrt{-\gamma}\left(\tilde T^{x^2x^2}\gamma_{ x^2x^2}+\tilde T^{x^2x^3}\gamma_{ x^2x^3}\right)\,,\nn 
&=-\frac{k}{2\pi}\int_0^{2\pi/k}dx_1\sqrt{-\gamma}\left(\tilde T^{x^3x^2}\gamma_{ x^3x^2}+\tilde T^{x^3x^3}\gamma_{ x^3x^3}\right) \,.
\end{align}
The first expression gives the first line of \eqref{eq:OSactionp}, while the second and third expressions give the second line. We can also check that the expression for the stress tensor 
satisfies the condition (2.18) of  \cite{Donos:2013cka}. Finally, the first law given in equation (2.13)
of \cite{Donos:2013cka} implies that
\begin{align}
\delta w=&-s\delta T+\left(8 c_h+2k^4\sinh^2 2\alpha_0(1+ 2 \cosh 4\alpha_0)\right)\frac{\delta k}{k}\nn
&-\frac{1}{4}\left( 64 c_\alpha + 3k^4(2\sinh 4\alpha_0-3\sinh 8\alpha_0)\right)\delta \alpha_0\,.
\end{align}

It is also illuminating to write the stress tensor components in the $x_2,x_3$ sector 
with respect to the basis of vectors dual
to the left-invariant one-forms $\omega_i$. Writing $T=T^{tt}v_t v_t+T^{\omega_i\omega_j}v_iv_j$ with $v_t=\partial_t$,
$v_{x_1}=\partial_{x_1}$, $v_2=\cos kx_1\partial_{x_2}-\sin k x_1 \partial_{x_3}$ and $v_3=\sin k x_1 \partial_{x_2}+\cos k x_1 \partial_{x_3}$
we obtain the diagonal components: 
\begin{align}\label{stressotherbasis}
T^{tt}&=3 M +8 c_h+\frac{k^4}{192}(1-e^{-4\adef})^2(61+70e^{4\adef}+61 e^{8\adef})\,,\nonumber\\
T^{\omega_1 \omega_1}&=M+8 c_h+\frac{k^4}{192}(1-e^{-4\adef})^2(95+98e^{4\adef}+95e^{8\adef})\,,\nonumber\\
T^{\omega_2 \omega_2}&=-\frac{1}{96} e^{-2 \adef} [3 (-32 M + k^4 - 256 c_\alpha) +   k^4 (-4 \cosh{4 \adef}+ \cosh{8 \adef} - 
      72 \sinh{4 \adef} + 108 \sinh{8 \adef})]\,, \nonumber\\
T^{\omega_3 \omega_3}&=-\frac{1}{96} e^{2 \adef} [3 (-32 M + k^4 + 256 c_\alpha) +   k^4 (-4 \cosh{4 \adef}+ \cosh{8 \adef} +72 \sinh{4 \adef} - 108 \sinh{8 \adef})]\,.
\end{align} 

We can also determine the anomalous scaling behaviour of the energy-momentum tensor.
Under the scaling transformations given in \eqref{anomscal}, we find that
\begin{align}
T^{\mu\nu}\to\lambda^4 T^{\mu\nu}+(\lambda k)^4\log\lambda h^{\mu\nu}\,,
\end{align}
where in the dual basis used in \eqref{stressotherbasis} we have
\begin{align}
h^{tt}&=-A/3\,,\qquad\qquad\qquad
h^{\omega_1 \omega_1}=-A\,,\nn
h^{\omega_2 \omega_2}&=e^{-2\adef}(A/3+B)\,, \qquad
h^{\omega_3 \omega_3}=e^{2\adef}(A/3-B)\,.
\end{align}
with $A\equiv \cosh 4 \adef-\cosh 8 \adef$ and $B\equiv  \frac{2}{3}(\sinh{4 \adef}-2 \sinh{8 \adef})$.
Notice that the tensor $h$ is traceless with respect to the boundary metric \eqref{bdymet}, consistent
with \eqref{tracebdy}.

\section{Derivatives of the $v_j$ with respect to the $s_i$}\label{altder}
In this appendix we will derive the expressions for the derivatives $\partial_{s_j}v_i$ given in
\eqref{lastone} that we used to obtain the Green's function. This provides a further development of the approach
described in \cite{Donos:2013eha}, which we will also describe in the next subsection.

To properly take into account gauge transformations in forming the derivatives we argue as follows.
We first consider a solution to the perturbed equations of motion \eqref{sceqs}, \eqref{constr2},
that is specified by the UV expansion parameters (appearing in \eqref{eq:UVexpper}) given
by $(s_1^r, s_2^r ,v_{1}^r ,v_{2}^r)$ and that satisfies in-going boundary conditions at the black hole horizon.
We next consider a pure gauge solution that is obtained by taking
$x_1\to x_1+e^{-i \omega t}\epsilon$ in the background solution. 
If $\epsilon$ is a constant, $\epsilon_0$, then this will preserve the gauge but violate the 
in-going boundary conditions at the black hole horizon. This can be remedied by taking $\epsilon$ to be a 
function of $r$ that vanishes at the horizon and approaches $\epsilon_0$
at the UV boundary with a suitably fast 
falloff in $r$. This latter condition will ensure that while this transformation will generate $h_{x_1r}$ terms in the perturbation, taking us outside our gauge, this will not have any additional impact on the UV data over and above that
given in \eqref{gtexp}. We can therefore paramatrise
a general class of solutions with parameters $(\zeta,\epsilon_0)$ via the UV data
\begin{align}\label{essesandvees}
s_{1}& = s_{1}^r \zeta + s_{1}^g \epsilon_0\,,\nn
s_{2} &= s_{2}^r \zeta + s_{2}^g \epsilon_0\,,\nn
v_{1} &= v_{1}^r \zeta + v_{1}^g \epsilon_0\,,\nn
v_{2}&= v_{2}^r \zeta + v_{2}^g \epsilon_0\,.
\end{align}
where 
\begin{align}\label{appsv}
s_1^g&=-i \omega,\quad\qquad\qquad v_{1}^g=-2i\omega(c_h+\frac{k^4}{8}\sinh^4{2\adef})\,, \nn
s_2^g &=-2k \sinh{2\adef},\quad v_{2}^g=-k \cosh 2\adef(4c_\alpha+k^4\cosh2\adef \sinh^32\adef)\,.
\end{align}
Next we observe that the first two equations in \eqref{essesandvees} imply
\begin{align}\label{zedzeta}
\zeta = \frac{s_{1}s_{2}^g-s_{2} s_{1}^g}{s_{1}^rs_{2}^g  -s_{2}^r  s_{1}^g},\qquad
\epsilon_0=\frac{s_{1}^rs_{2}-s_{2}^r s_{1} }{s_{1}^rs_{2}^g -s_{2}^rs_{1}^g}\,.
\end{align}
We can also obtain analogous expressions using the second two equations and equating these with 
\eqref{zedzeta} we obtain the relations
\begin{align}\label{crsols}
v_{1}^r &=\frac{v_{1}^g s_{2}^r s_{1}-v_{1} s_{2}^r s_{1}^g-v_{1}^g s_{2} s_{1}^r+v_{1} s_{2}^g s_{1}^r}{s_{1}s_{2}^g -s_{2} s_{1}^g}\,,\nn
v_{2}^r&=\frac{v_{2}^g s_{2}^r s_{1}-v_{2} s_{3}^r s_{1}^g-v_{2}^g s_{2} s_{1}^r+v_{2} s_{2}^g s_{1}^r}{s_{1}s_{2}^g -s_{2} s_{1}^g}\,.
\end{align}

We next calculate 
\begin{align}\label{firstgoatder}
\partial_{s_{i}}v_{j}&=v_{j}^r \partial_{s_{i}}\zeta + v_{j}^g \partial_{s_{i}}\epsilon_0\,,
\end{align}
and then using \eqref{zedzeta} we obtain
\begin{align}\label{ginvders}
\partial_{s_{1}}v_{1}&=\frac{v_{1}^r s_{2}^g-s_{2}^rv_{1}^g }{s_{1}^rs_{2}^g -s_{2}^r s_{1}^g}\,,\nn
\partial_{s_{2}}v_{1}&=\frac{s_{1}^rv_{1}^g -v_{1}^r s_{1}^g}{ s_{1}^rs_{2}^g-s_{2}^r s_{1}^g}\,,\nn
\partial_{s_{1}}v_{2}&=\frac{v_{2}^r s_{2}^g -s_{2}^rv_{2}^g}{ s_{1}^rs_{2}^g-s_{2}^r s_{1}^g}\,,\nn
\partial_{s_{2}}v_{2}&=\frac{ s_{1}^rv_{2}^g-v_{2}^r s_{1}^g}{ s_{1}^rs_{2}^g-s_{2}^r s_{1}^g}\,.
\end{align}
Note that the same procedure as in \eqref{firstgoatder} gives the expected $\partial_{s_i}s_j=\delta^i_j$. 
It is important to emphasise that the results in \eqref{ginvders} are gauge invariant in the sense that they are unchanged under the shift of $(s_{1}^r, s_{2}^r ,v_{1}^r ,v_{2}^r)$ by an arbitrary amount of $(s_{1}^g, s_{2}^g ,v_{1}^g ,v_{2}^g)$. 
Consistent with this, using \eqref{crsols}, the quantities with an $r$ superscript in \eqref{ginvders} can be replaced by
those without. After substituting expressions \eqref{appsv} into \eqref{ginvders}, 
we obtain the results quoted in the main text \eqref{lastone}.

\subsection{The approach of \cite{Donos:2013eha}}
We briefly comment on the approach for obtaining the Green's function from the currents, which was used in \cite{Donos:2013eha} and also in \cite{Donos:2014yya}. The basic idea is to calculate the components $G_{i1}$ by working in a gauge 
$s_2=0$ via
\begin{align}\label{gone}
G_{i1}=\left.\frac{J_i}{s_1}\right\vert_{s_2=0}\,,
\end{align}
and similarly the components $G_{i2}$ by working in a gauge $s_1=0$ via
\begin{align}\label{gtwo}
G_{i2}=\left.\frac{J_i}{s_2}\right\vert_{s_1=0}\,.
\end{align}

Let us first consider the gauge $s_2=0$. 
If we have a solution satisfying the in-falling boundary conditions with UV data given by $(s_i^r,v_i^r)$ then we can consider a gauge transformation $x_1\to x_1+e^{-i \omega t}\epsilon(r)$, with $\epsilon(r)$ approaching the constant $\epsilon_0$ at the UV boundary, with additional properties as described earlier in this appendix. 
If we choose $\epsilon_0= \frac{s_2^r}{2 k \sinh{2\adef}}$ we obtain (setting $\zeta=1$ in \eqref{essesandvees})
\begin{align}
\label{eq:gauge}
(s_1,s_2) &= (s_1^r-\frac{i\omega}{2 k \sinh{2\adef}}s_2^r, 0)\,,\nonumber\\
v_1&=v_{1}^r-\frac{i\omega }{2k\sinh{2\adef}}(2c_h+\frac{ k^4}{4}\sinh^4{2\adef})s_2^r,\nn
v_2&=v_2^r- \coth{2 \adef}( 2 c_\alpha +\frac{k^4}{2} \cosh2\adef\sinh^3{2\adef})s_2^r\,,
\end{align}
and we conclude that in this gauge we have
\begin{align}\label{lastonetest}
&\left.\frac{v_1}{s_1}\right\vert_{s_2=0} =\frac{(2 k \sinh{2 \adef})v_1^r-i\omega (2 c_h + \tfrac {k^4} {4}\sinh^4 {2\adef})s_2^r }{2 k \sinh{2 \adef} s_1^r-i\omega s_2^r }\,,\nonumber\\
&\left.\frac{v_2}{s_1}\right\vert_{s_2=0}=\frac{(2 k \sinh{2 \adef})v_2^r
-k \cosh {2 \adef} (4  c_\alpha+ \, k^4 \cosh {2 \adef} \sinh^3 {2 \adef})s_2^r }{2 k \sinh{2 \adef} s_1^r-i \omega s_2^r }\,,
\end{align}
which combined with \eqref{gone} and \eqref{eq:J1J21} will give the same result for $G_{i1}$ as in \eqref{lastone}.

Alternatively, one can achieve $s_1=0$ by performing a gauge transformation with $\epsilon_0=\frac{- i s_1^r}{\omega}$, so that 
 \begin{align}
\label{eq:gauge2}
(s_1,s_2) &= (0, s_2^r+\frac{2ik \sinh{2\adef}}{\omega}s_1^r)\,,\nonumber\\
v_1&=v_{1}^r-(2c_h +\frac{k^4}{4} \sinh^4{2\adef})s_1^r,\nn
v_2&=v_2^r+\frac{ik}{\omega}(4c_\alpha \cosh{2 \adef}  +k^4 \cosh^2{2 \adef}\sinh^3{2 \adef} )s_1^r\,,
\end{align} 
and hence
\begin{align}\label{lastonetest2}
&\left.\frac{v_{1}}{s_2}\right\vert_{s_1=0}=\frac{i\omega (2 c_h + \tfrac {k^4} {4}\sinh^4 {2\adef}) s_1^r-i \omega v_1^r}{2 k \sinh{2 \adef} s_1^r-i\omega s_2^r}\,,\nonumber\\
&\left.\frac{v_{2}}{s_2}\right\vert_{s_1=0}=\frac{k \cosh {2 \adef} (4  c_\alpha +  \, k^4 \cosh {2 \adef} \sinh^3 {2 \adef})s_1^r-i\omega v_2^r}{2 k \sinh{2 \adef} s_1^r-i\omega s_2^r }\,.
\end{align}
Combining this with \eqref{gone} and \eqref{eq:J1J21} will give the same result for $G_{i2}$ as in \eqref{lastone}.

\section{Derivatives of the on-shell action and the relationship to the Green's function} 
As emphasised in \cite{Son:2002sd} evaluating the on-shell action and then taking two derivatives with respect to the sources should give a real quantity. Thus, despite some claims to the contrary in the literature,
the evaluated on-shell action does not provide a method to obtain the Green's function directly. 
In this appendix, we investigate this in a little more detail as it provides a nice consistency check on the procedures we have used.

The on-shell  Minkowski action at second order in the perturbation can be written in the form
\begin{align}\label{eq:S2}
 S^{(2)}=&\int d rd^2{x}\frac{d\omega}{2\pi}\Big(\frac{r^2} {2 f h}(\frac{g'}{g}+2\frac{f'}{f}+2\frac{h'}{h})h_{t x_1}^2-\frac{3 r^2}{2 f h}h_{t x_1} h_{t x_1}'\nn
 &\qquad\qquad\qquad\qquad\qquad-\frac{2 g h f}{ r^3}h_{23}^2+\frac{3 fgh}{2 r^2} h_{23} h_{23}'\Big)' +c.t.+log
\end{align}
where, for ease of presentation, 
we have not written out the contributions from the counterterms 
and log terms (the Minkowski analogues of \eqref{ctermp}, \eqref{ctermlog})
and e.g. $h_{t x_1}^2=h_{t x_1}(\omega) h_{t x_1}(-\omega)$. To get this expression we have
used the second-order equations for the perturbation as well as the background equations of motion and carried out the integral over time. In particular, there are some total time derivatives 
which give no contribution.
We next observe that the total derivative in $r$ picks up contributions from the 
UV boundary and potentially the black hole horizon. In fact since both $g(r_+)$ and $h_{t x_1}(r_+)$ vanish there is only a contribution from the horizon from the last term, but this vanishes after integrating over all $\omega$ 
(this is in contrast to statements made in \cite{Son:2002sd}).

Thus, using the UV expansions for the background, \eqref{eq:expUV}, 
and the perturbation, \eqref{eq:UVexpper}, along with the constraint \eqref{eq:constrUV}, we find that
the on-shell action \eqref{eq:S2} can be written as
\begin{align}\label{sonshellact1}
  S^{(2)}_\infty=&\int d^2 x\int_{\omega\ge 0} \frac{d\omega}{2\pi}\Big(
  \frac{s_2\bar s_2}{96}(-148 k^4 \cosh4 \adef-131 k^4 \cosh 8 \adef+108 k^2 \omega ^2 \cosh 4 \adef\nn&\qquad\qquad\qquad\qquad-9 k^4+36 k^2 \omega ^2-96 M-18
   \omega ^4)\nonumber\\
   + &\frac{s_1 \bar s_1}{96}(-44 k^4 \cosh 4 \adef+35 k^4 \cosh 8 \adef-36 k^2 \omega ^2 \cosh 4 \adef+9 k^4+36 k^2 \omega ^2-288 M)\nonumber\\
  + & \frac{s_1 \bar s_2}{8}i\omega k(-3  \omega^2+ 8 k^2  +16 k^2\cosh 4 \adef)\sinh 2\adef\nn
  - & \frac{\bar s_1  s_2}{8}i\omega k(-3  \omega^2+ 8 k^2  +16 k^2\cosh 4 \adef)\sinh 2\adef
  +2(s_2 \bar v_2+\bar s_2 v_2-s_1\bar v_1-\bar s_1 v_1)\Big)\,.
\end{align}
As in the main text we are treating $s_i=s_i(\omega)$ and $\bar s_i=\bar s_i(\omega)$ as independent
variables with $\omega>0$ and similarly with the expectation values $v_i$ and $\bar v_i$, which are to be considered as functions of the sources: $v_i=v_i(s_1,s_2)$ and $\bar v_i=\bar v_i(\bar s_1,\bar s_2)$.

Using the derivatives given in \eqref{lastone} and also the constraint \eqref{eq:constrUV} 
we now find, after some calculation, the simple result
\begin{align}
\frac{\partial^2 S^{(2)}}{\partial s_i\partial \bar s_j}=G_{ij}+G^\dagger_{ij}\,,
\end{align}
with $G_{ij}$ as given in \eqref{eq:Gij}.

\bibliographystyle{utphys}
\bibliography{helical}{}
\end{document}